\documentstyle[aps,pre,multicol,epsfig]{revtex}
\begin{document}
\title{Mound formation and coarsening from a
nonlinear instability in surface growth}
\author{Buddhapriya Chakrabarti and Chandan Dasgupta \cite{cd}}
\address{Centre for Condensed Matter Theory, Department of Physics, 
Indian Institute of Science, Bangalore 560012, INDIA\\
and\\
Condensed Matter Theory Unit, JNCASR, Bangalore 560064, INDIA}
\date{\today}
\maketitle
\draft
\begin{abstract} 
We study a class of one-dimensional, nonequilibrium, conserved growth 
equations for both nonconserved and  
conserved noise statistics using numerical integration. An atomistic
version of these growth equations is also studied using stochastic
simulation. The models with nonconserved noise statistics 
are found to exhibit mound formation and power-law coarsening with slope 
selection for a range of values of the model parameters. 
Unlike previously proposed models of mound formation, the Ehrlich-Schwoebel 
step-edge barrier, usually modeled as a linear instability in growth 
equations, is absent in our models. 
Mound formation in our models occurs due to a nonlinear instability in which 
the height (depth) of spontaneously generated pillars (grooves) increases
rapidly if the initial height (depth) is sufficiently large.
When this instability is controlled by the introduction of an infinite number 
of higher-order gradient nonlinearities, the system exhibits a
first-order dynamical phase transition from a rough self-affine phase to
a mounded one as the value of the parameter that measures the
effectiveness of control is decreased.
We define a new ``order parameter'' that may be used to distinguish
between these two phases. In the mounded phase, the system exhibits power-law  
coarsening of the mounds in which a selected slope is retained at 
all times. The coarsening 
exponents for the continuum equation and the discrete model are found to
be different. An explanation of this difference is proposed and verified
by simulations. In the growth equation with conserved noise, we find the 
curious result that the
kinetically rough and mounded phases are both locally stable 
in a region of parameter space. In this region,
the initial configuration of the system determines its steady-state 
behavior. 
\end{abstract}
\begin{multicols}{2} 
\narrowtext
\section{Introduction}
\label{intro}
The process of growing films by the deposition of atoms on a 
substrate is of considerable experimental and theoretical 
interest~\cite{rev1}. While there has been a lot of research on the 
process of kinetic roughening~\cite{rev1,rev2,rev3} leading to a 
self-affine interface profile, there has been much recent 
experimental~\cite{johnson,thurmer,stros,tsui,zuo,gyure,aposto,zhao,lengel} 
and theoretical~\cite{johnson,stros,sp,rk,siegert,golub,pierre,ramana,patcha1,biehl,krug2,ev,pv,patcha2,patcha3} interest in a different mode of
surface growth involving the formation of ``mounds'' which are
pyramid-like or ``wedding-cake-like'' structures.  The precise 
experimental conditions that determine whether the growth morphology
would be kinetically rough or dominated by mounds are presently
unclear. However, many experiments show the formation of mounds that
{\it coarsen} (the typical lateral size of the mounds increases) with
time. During 
this process, the typical slope of the sides of the pyramid-like mounds may
or may not remain constant. If the slope remains constant in time, the
system is said to exhibit {\it slope selection}. As the mounds coarsen,
the surface roughness characterized by the root-mean-square width of the
interface increases. Eventually, at very long times, the system is expected to
evolve to a single-mound structure in which the mound size is equal to
the system size. 

There are obvious differences between the structures of
kinetically rough and mounded interfaces. In the first case, the
interface is rough in a self-affine way at length scales shorter than
a characteristic length $\xi(t)$ that initially increases with time and
eventually saturates at a value comparable to the sample size. In the second
case, the characteristic length is the typical mound size $R(t)$ whose
time-dependence is qualitatively similar to that of $\xi(t)$. However,
the interface in this case looks well-ordered at
length scales shorter than $R(t)$. Nevertheless, there are
certain similarities between the gross features of these two kinds of
surface growth. First consider the simpler situation in which the slope
of the sides of the mounds remains constant in time.  Simple geometry 
tells us that if the system evolves to
a single-mound structure at long times, then the ``roughness exponent''
$\alpha$ must be equal to unity. Also, the height-difference
correlation function $g(r)$ is expected to be proportional to $r$ for
$r\ll R(t)$. This is consistent with $\alpha = 1$. If the mound size
$R(t)$ increases with time as a power law, $R(t) \sim t^n$, during
coarsening, then the interface width $W$, which is essentially the
height of a typical mound, should also increase with time as a power
law with the same exponent $n$. Thus, dynamic scaling with ``growth
exponent'' $\beta$ equal to $n$, and ``dynamical exponent'' $z$ equal
$1/n$ is recovered. If the
mound slope $s(t)$ increases with time as a power law, $s(t) \sim
t^\theta$ (this is known in the literature as {\it steepening}), then 
one obtains behavior similar to anomalous dynamical scaling~\cite{anom}
with $\beta = n+\theta$, $z=1/n$. 

These similarities between the gross scaling properties of kinetic
roughening with a large value of $\alpha$  and mound  
formation with power-law coarsening make it difficult to experimentally  
distinguish between these two modes of surface growth. This poses a
problem in the interpretation of experimental results~\cite{zhao,lengel}. 
Existing experiments on mound formation show a wide variety of behavior.
Without going into the details of individual experiments, we note 
that some experiments show mound coarsening with a time-independent
``magic'' slope, whereas other experiments do not show any
slope selection. The detailed morphology of the mounds varies
substantially from one experiment to another. 
The reported values of the coarsening exponent $n$ show a large variation
in the range 0.15-0.4. 

Traditionally, the formation of mounds has been
attributed to the presence of the so-called Ehrlich-Schwoebel (ES)
step-edge barrier~\cite{es1,es2} that hinders the downward motion of
atoms across the edge of a step. This step-edge diffusion bias makes it
more likely for an atom diffusing on a terrace to attach to an
ascending step than to a descending one.  This leads to an effective
``uphill'' surface current~\cite{villain} that has a destabilizing
effect, leading to 
the formation of mounded structures as the atoms on upper
terraces are prevented by the ES barrier from coming down. 

This destabilizing effect is usually represented
in continuum growth equations by a {\it linear instability}. Such growth
equations usually have a ``conserved'' form in which the time-derivative of
the height is assumed to be equal to the negative of the divergence of 
a nonequilibrium surface
current $\bf j$. The effects of an ES barrier are modeled in these
equations by a term in $\bf j$ that is proportional to the gradient of the
height (for small values of the gradient) 
with a {\it positive} proportionality constant.
Such a term is manifestly unstable,
leading to unlimited exponential growth of the ${\bf k} \ne 0$ Fourier 
components of the height. This instability
has to be controlled by other nonlinear terms in the growth equation
in order to obtain a proper description of the long-time behavior.
A number of different choices for the nonlinear terms have been reported
in the literature \cite{johnson,stros,sp,rk,ev,pv}. 
If the ``ES part'' of $\bf j$ has one or more
stable zeros as a function of the slope $\bf s$, then 
the slope of the mounds that form as a result of the ES instability 
is expected to stabilize at the corresponding 
value(s) of $\bf s$ at long times. The system would then exhibit 
slope selection. If, on the other
hand, this part of $\bf j$ does not have a stable zero, then the
mounds are expected to continue to steepen with time.
Analytic and numerical studies of such
continuum growth equations have produced a wide variety of results, such
as power-law coarsening and slope selection with  $n=1/4$~\cite{sp} 
or $n \simeq 0.17$~\cite{stros} in
two dimensions, power-law coarsening accompanied by a steepening of the 
mounds~\cite{johnson,rk,golub}, and a complex coarsening process~\cite{ev,pv} 
in which the growth of the mound size 
becomes extremely slow after a characteristic size is reached.  

There are several atomistic, cellular-automaton-type  
models~\cite{patcha1,biehl,krug2} that incorporate the
effects of an ES diffusion barrier. Formation and 
coarsening of mounds in the presence of an ES barrier have also been
studied~\cite{ev,pv} in a 1d model with both discrete and continuum
features. We also note that a new mechanism for mounding 
instability has been discovered recently \cite{pierre,ramana}. This
instability, generated by fast diffusion along the edges of monatomic
steps, leads to the formation of quasi-regular shaped mounds in two or
higher dimensions. The effects of this instability have been studied 
in simulations~\cite{pierre,ramana,biehl,patcha2,patcha3}. 
The wide variety of results~\cite{pierre,ramana,patcha1,biehl,krug2,ev,pv,patcha2,patcha3} 
obtained from simulations of different models, combined with similar
variations in the experimental results, have made it very difficult
to identify the microscopic mechanism of mound formation in surface growth.

In this paper, we show that mound formation, 
slope selection and power-law coarsening in a class of one-dimensional (1d)
continuum growth
equations and discrete atomistic models can occur from a mechanism that is 
radically different from the ones mentioned above. 
Our study is based on the conserved nonlinear growth equation proposed
by Villain~\cite{villain} and by Lai and Das Sarma~\cite{lds}, and an
atomistic version~\cite{kds} of this equation. We have studied the
behavior of the continuum equation by numerical integration, and the
behavior of the atomistic model by stochastic simulation.
Previous work~\cite{us1,us2,kund} on these systems showed that 
they exhibit a {\it nonlinear} instability, in which pillars (grooves) 
with height (depth) greater  than 
a critical value continue to grow rapidly. This instability can be
controlled~\cite{us1,us2,kund} by 
the introduction of an infinite number of higher-order gradient 
nonlinearities. When the parameter that describes the effectiveness of
control is sufficiently large, the controlled 
models exhibit~\cite{us1,us2,kund} kinetic roughening, 
characterized by usual dynamical scaling with exponent
values close to those expected from dynamical renormalization group
calculations~\cite{lds}. As the value of the control parameter is decreased,
these models exhibit transient multiscaling~\cite{us1,us2,kund} of height
fluctuations. For yet smaller values of the control parameter, the rapid
growth of pillars or grooves causes a breakdown of dynamical scaling,
with the width versus time plot showing a sharp upward deviation~\cite{us2} from
the power-law behavior found at short times (before the onset of the
nonlinear instability).

We report here the results of a detailed study of the behavior of these models
in the regime of small values of the control parameter where conventional
kinetic roughening is not observed. We find that in this regime, the 
interface self-organizes into a sawtooth-like structure with a series of
triangular, pyramid-like mounds. These mounds coarsen in time, with 
larger mounds growing at the expense of smaller ones. In this coarsening
regime, a power-law dependence of the interface width on time is recovered.
The slope of the mounds remains constant during the coarsening process.
In section~\ref{models}, the growth equation and the atomistic model 
studied in this work are defined and the numerical methods we have used to
analyze their behavior are described. The basic phenomenology of mound
formation and slope selection in these systems is described in detail in
section~\ref{mound}. Specifically, we show that the nonlinear mechanism of
mound formation in these systems is ``generic'' in the sense that the
qualitative behavior does not depend on the specific form of the function 
used for controlling the instability. In particular, we find very similar
behavior for two different forms of the control function: one used in
earlier studies~\cite{us1,us2,kund} of these systems, and the other one
proposed by Politi and Villain~\cite{pv} from physical considerations. 
Since the linear instability used conventionally to model the ES mechanism
is explicitly absent in our models, our work shows that the presence of
step-edge barriers is not essential for mound formation. The slope selection
found in our models is a true example of nonlinear pattern formation: since
the nonequilibrium surface current in our models vanishes for all values
of constant slope, the selected value of the slope can not be predicted in
any simple way. This is in contrast to the behavior of ES-type models
where slope selection occurs only if the surface current vanishes at a 
specific value of the slope.

Next, in section~\ref{pt}, we show that the change in the dynamical behavior
of the system (from kinetic roughening to mound formation and coarsening)
may be described as a first-order nonequilibrium phase transition. Since
the instability in our models is a nonlinear one, the flat interface is
locally stable in the absence of noise for all values of the model
parameters (the strength of 
the nonlinearity and the value of the control parameter). The mounded phase
corresponds to a different stationary solution of the dynamical equations
in the absence of noise. We use a linear stability analysis to find the
``spinodal boundary'' in the two-dimensional parameter space across which
the mounded stationary solution becomes locally unstable. We show that
the results of
this numerical stability analysis can also be obtained from simple analytic
arguments. To obtain the phase boundary in the presence of noise, we first
define an {\it order parameter} that is zero in the kinetically rough phase
and nonzero in the mounded phase. We combine the numerically obtained
results for this order parameter for different sample sizes with finite-size
scaling to confirm that this order parameter exhibits the expected
behavior in the two phases. The  phase boundary that 
separates the mounded phase from the kinetically rough one is obtained
numerically. The phase
boundaries for the continuum model with two different forms of the control
function and the atomistic model are found to be qualitatively similar.

The results of a detailed study of the process of coarsening of the mounds
are reported in section~\ref{coars}.
Surprisingly, we find that the coarsening exponents of the continuum equation  
and its atomistic version are different. We propose a possible explanation of
this result on the basis of an analysis 
of the coarsening process in which the problem is mapped to that of 
of a Brownian walker in an attractive force field. In this mapping,
the Brownian walk is supposed to describe the noise-induced 
random motion of the peak of a mound, and the  
attractive ``force'' represents the interaction between neighboring
mounds that leads to
coarsening. We show that the numerical results obtained for the dynamics
of mounds in the atomistic model are consistent with this explanation.

In section~\ref{conserv},
we consider the behavior of the continuum growth equation for ``conserved''
noise statistics. The nonlinear instability found in the nonconserved case
is expected to be present in the conserved case also. However, there is an
important difference between the two cases. The nonconserved model exhibits
anomalous dynamical scaling, so that the typical nearest-neighbor height 
difference continues to increase with time, and the instability is always
reached~\cite{us2} at sufficiently long times, even if the starting 
configuration is perfectly flat. 
Since the continuum model with conserved noise 
statistics exhibits~\cite{sgg} usual dynamic scaling with $\alpha < 1$,
the nearest-neighbor height difference is expected to saturate at 
long times if the initial state of the system is flat. 
Under these circumstances,
the occurrence of the nonlinear instability in runs started from flat states
would depend on the values of the parameters. Specifically, the instability
may not occur at all if the value of the nonlinear coefficient in the growth
equation is sufficiently small. At the same time, the instability can be
initiated by choosing an initial state with sufficiently high (deep) pillars
(grooves). Since mound formation in these models is crucially dependent 
on the occurrence of the instability, the arguments above suggest that 
the nature of the 
long-time steady state reached in the conserved model may depend
on the choice of the initial state. Indeed, we find from simulations that
in a region 
of parameter space,  the mounded and kinetically rough phases are both 
locally stable and the steady state configuration is determined by the 
choice of the initial configuration of the interface. These 
results imply the surprising conclusion that the long-time, steady-state
morphology 
of a growing interface, as well as the dynamics of the process by which the
steady state is reached may be 
``history dependent''  in the sense that the behavior would 
depend crucially on the choice of the initial state. 
A summary of our findings and a discussion of the implications of 
our results are provided in Sec.\ref{summ}. A summary of the basic results of 
our study was reported in a recent Letter~\cite{bcdg}.

\section{Models and methods}
\label{models}

Conserved growth equations (deterministic part of the dynamics 
having zero time derivative for the ${\bf k}=0$ Fourier mode of the height 
variable) with nonconserved noise are generally used~\cite{rev2} to model 
nonequilibrium surface growth in molecular beam epitaxy (MBE). 
The conservation is a consequence of absence of bulk vacancies, overhangs 
and desorption (evaporation of atoms from the substrate) under optimum 
MBE growth conditions. Thus, integrating over the whole sample area gives 
the number of particles deposited. This conservation is not strictly 
valid because of ``shot noise'' fluctuations in the beam. The shot noise 
is modeled by an additive noise term $\eta({\bf r},t)$ in the 
equation of motion 
of the interface. The noise $\eta$ is generally assumed to
be delta-correlated in both space and time:
\begin{equation}
\langle \eta({\bf r}, t) \eta({\bf r}^{\prime}, t^{\prime}) \rangle = 
2D\delta^{d}({\bf r}-{\bf r}^\prime)\delta(t-t^\prime),
\end{equation}
where ${\bf r}$ is a point on a $d$-dimensional substrate.
Thus, a conserved growth equation may be written in a form 
\begin{equation}
\frac{\partial h}{\partial t}= - {\bf{\nabla}} \cdot {\bf j} + \eta , 
\label{mbe}
\end{equation}
where $h({\bf r},t)$ is the height at point $\bf r$ at time $t$, and 
${\bf j}$ is the surface current density.
The surface current models the deterministic dynamics at the 
growth front. As mentioned in section~\ref{intro}, 
the presence of an ES step-edge barrier
is modeled in  continuum equations of 
the form of Eq.(\ref{mbe}) by a term in $\bf j$ that is proportional to
the slope ${\bf s} = {\bf \nabla}h$, with a positive constant of
proportionality. This makes the flat surface ($h({\bf r})$ constant for
all ${\bf r}$) linearly unstable. This instability is controlled by the  
introduction of terms involving higher powers of the local slope $\bf s$
and higher-order spatial derivatives of $h$. 

We consider the conserved growth equation proposed by Villain~\cite{villain} 
and Lai and Das Sarma~\cite{lds} for describing MBE-type surface growth 
{\em in the absence of ES step-edge barriers}. This equation is of the form 
\begin{equation}
\partial h^{\prime}({\bf r},t^\prime)/\partial t^\prime = -\nu \nabla^4
h^{\prime} + \lambda^\prime \nabla^2 |{\bf \nabla}h^{\prime}|^2 +
\eta^\prime({\bf r},t^\prime), \label{lds1}
\end{equation}
where $h^{\prime}({\bf r},t^\prime)$ represents the height variable 
at the point ${\bf r}$ at time $t^\prime$.
This equation is believed~\cite{rev2} to provide a correct description
of the kinetic roughening behavior observed in MBE-type experiments
~\cite{lengel}. 

In our study, we numerically integrate the 1d version of 
Eq.(\ref{lds1}) using a simple 
Euler scheme~\cite{us2}. Upon choosing appropriate units of length 
and time and discretizing in space and time,  Eq.(\ref{lds1}) is 
written as~\cite{us2} 
\begin{eqnarray}
h_i(t + \Delta t) - h_i(t) &=& \Delta
t \tilde{\nabla}^{2} [ - \tilde{\nabla}^{2} h_i(
t) +
\lambda |\tilde{\nabla} h_i(t)|^2] \nonumber \\
&+&\sqrt{\Delta t}\, \eta_i(t), \label{lds2}
\end{eqnarray}
where $h_i(t)$ represents the dimensionless height variable at the lattice 
point $i$ at dimensionless time $t$, $\tilde{\nabla}$ and
$\tilde{\nabla}^{2}$ are lattice versions of the derivative
and Laplacian operators, and $\eta_i(t)$ is a random
variable with zero average and variance equal to unity.
These equations, with an appropriate choice of $\Delta t$,
are used to numerically follow the time evolution of the interface.
In most of our studies, we have used the following definitions for the
lattice derivatives:
\begin{eqnarray}
\tilde{\nabla} h_i &=& (h_{i+1} - h_{i-1})/2, \nonumber \\
\tilde{\nabla}^2 h_i &=& h_{i+1}+h_{i-1}-2h_i. \label{deriv}
\end{eqnarray}
We have checked that the use of more accurate, left-right symmetric 
definitions of the lattice derivatives, involving more neighbors to the left
and to the right~\cite{us2}, leads to results that are 
very similar to those obtained from calculations in which these simple
definitions are used. We have also checked that the results obtained in the
deterministic limit ($\eta = 0$) by using a more sophisticated integration
routine~\cite{dgeer} closely match those obtained from the Euler method with
sufficiently small values of the integration time step.

We have also studied an atomistic version~\cite{kds} of Eq.(\ref{lds1})
in which the height variables $\{ h_i \}$ are integers.
This model is defined by the following deposition
rule. First, a site (say $i$) is chosen at random. Then the
quantity
\begin{equation}
K_i(\{ h_j \}) =
-\tilde{\nabla}^{2}h_i + \lambda |\tilde{\nabla}h_i|^2
\label{kds1}
\end{equation}
is calculated for the site $i$ and all its nearest neighbors. Then, a
particle is added to the site that has the smallest value of $K$ among
the site $i$ and its nearest neighbors. In the case of a tie for the
smallest value, the site $i$ is chosen if it is involved in the tie;
otherwise, one of the sites involved in the tie is chosen randomly.
The number of deposited layers provides a measure of time in this model.

It was found in earlier studies~\cite{us1,us2,kund} that both these 
models exhibit a {\it
nonlinear instability} in which isolated structures (pillars for
$\lambda >0$, grooves for $\lambda<0$) grow rapidly if
their height (depth) exceeds a critical value. This instability can be
controlled~\cite{us1,us2,kund} by replacing $|\tilde{\nabla} h_i|^2$ in
Eqns.(\ref{lds2}) and (\ref{kds1}) by $f(|\tilde{\nabla} h_i|^2)$ where 
the nonlinear function $f(x)$ is defined as
\begin{equation}
f(x) = \frac{1-e^{-cx}}{c},
\label{fofx}
\end{equation}
$c>0$ being a 
control parameter. We call the resulting models ``model I'' and 
``model II'', respectively. This replacement, amounts to the 
introduction of an infinite series of higher-order nonlinear terms.
The time evolution of the height variables in model I is, thus, given by
\begin{eqnarray}
h_i(t + \Delta t) &-& h_i(t) = \Delta
t \tilde{\nabla}^{2} [ - \tilde{\nabla}^{2} h_i(t) \nonumber \\
&+& \lambda (1-e^{-c|\tilde{\nabla} h_i(t)|^2})/c]
+\sqrt{\Delta t}\, \eta_i(t). \label{cld}
\end{eqnarray}
In model II, the quantity $K_i$ is defined as
\begin{equation}
K_i(\{ h_j \}) =
-\tilde{\nabla}^{2}h_i + \lambda (1-e^{-c|\tilde{\nabla}h_i|^2})/c.
\label{model2}
\end{equation}

While the function $f(x)$ was introduced in the earlier work 
purely for the purpose of controlling the nonlinear instability, it turns
out that the introduction of this function in the growth equation is
physically meaningful.
Politi and Villain~\cite{pv} have shown that the nonequilibrium
surface current that leads to the $\nabla^2 |{\bf \nabla}h^{\prime}|^2$
term in Eq.(\ref{lds1}) should be proportional to
${\bf \nabla}|{\bf \nabla} h^\prime|^2$ when $|{\bf \nabla} h^\prime|$ 
is small, and should go to zero when $|{\bf \nabla} h^\prime|$ is
large. The introduction of the ``control function'' 
$f(|\tilde{\nabla} h_i|^2)$ satisfies this physical requirement. We have
also carried out studies of a slightly different model (which we call
``model IA'') in which the function $f(x)$ is assumed to have a form suggested
by Politi and Villain:
\begin{equation}
f(x)=\frac{x}{1+cx},
\label{pvform}
\end{equation}
where $c$ is, as before, a positive control parameter. 
This function has the same
asymptotic behavior as that of the function defined in Eq.(\ref{fofx}). As
we shall show later, the results obtained from calculations in which these
two different forms of $f(x)$ are used are qualitatively very similar. In
fact, we expect that the qualitative behavior of these models would be the
same for any monotonic function 
$f(x)$ that satisfies the following requirements:
(i) $f(x)$ must be proportional to $x$ in the small-$x$ limit; and (ii) it
must saturate to a constant value as $x \to \infty$.

We have carried out extensive simulations of both these models for 
different system sizes. The results reported here have been 
obtained for systems of sizes $40 \le L \le 1000$. There is 
no significant dependence of the results on $L$. The time step used 
in most of our studies of models I and IA is $\Delta t = 0.01$. We have 
checked that very similar results are obtained for smaller values 
of $\Delta t$. We used both unbounded (Gaussian) and bounded distributions
for the random variables $\eta_i$ in our simulations of models I and IA,
with no significant difference in the results. For computational convenience,
a bounded distribution (uniform between $+\sqrt{3}$ and $-\sqrt{3}$) was
used in most of our calculations. Unless otherwise stated, the results
described in the following sections were obtained using periodic
boundary conditions. The effects of using other boundary conditions will be
discussed in the next section.

\section{Mound formation and Slope Selection}
\label{mound}

It has been demonstrated earlier~\cite{us1,us2} that 
if the control parameter $c$ is sufficiently large,
then the nonlinear instability is completely suppressed and the models
exhibit the usual dynamical scaling behavior with the
expected~\cite{lds} exponent values, $\beta \simeq 1/3$, 
$z \simeq 3$, and $\alpha = \beta z \simeq 1$. 
This behavior for model I is illustrated by the solid line in 
Fig.\ref{bigfig1}, which shows a plot of the width $W$ as a function of time 
$t$ for parameter values $\lambda$=$4.0$ and $c=0.06$. 
As the value of $c$ is decreased 
with $\lambda$ held constant, the instability makes its appearance: 
the height $h_0$ of an isolated pillar (for $\lambda >0$) increases in 
time if $h_{min}(\lambda,c) < h_0 < h_{max}(\lambda,c)$. The value of 
$h_{min}$ is nearly independent of $c$, while $h_{max}$ increases as 
$c$ is decreased~\cite{us2}. 
If $c$ is sufficiently large, $h_{max}$ is small and 
the instability does not affect the scaling behavior of global quantities 
such as $W$,  
although transient multiscaling at length scales shorter than the
correlation length $\xi \sim t^{1/z}$ may be found~\cite{us1,us2} if $c$
is not very large.
As $c$ is decreased further, $h_{max}$ becomes large, and when
isolated pillars with $h_0 >h_{min}$ are created at 
an initially flat interface through fluctuations, the rapid
growth of such pillars to height $h_{max}$ leads to a sharp upward
departure from the power-law scaling of $W$ with time $t$.
The time at which this departure occurs varies from run to run. This
behavior for model I with $\lambda=4.0$ and $c=0.02$ is shown by the 
dash-dotted line in Fig.\ref{bigfig1}.

This instability leads to the formation of a large number of 
randomly distributed pillars of
height close to $h_{max}$. As the system evolves in time, the
interface self-organizes to form triangular mounds of a fixed slope
near these pillars. These mounds then coarsen in time, with large
mounds growing larger at the expense of small ones. In this coarsening
regime, a power-law growth of $W$ in time is recovered.
The slope of the sides of the triangular mounds remains
constant during this process. Finally, the system reaches a steady
state with one peak and one trough (if periodic boundary conditions are 
used) and remains in this state for longer
times. The interface profiles in the kinetically rough phase (obtained 
for relatively large values of $c$) and the mounded phase (obtained for
small $c$) are qualitatively different. This difference is illustrated
in Fig.\ref{bigfig2} that shows typical interface profiles in the two
different phases. This figure also shows a typical interface profile
for model IA in the mounded regime, illustrating the fact that the
precise choice of the control function $f(x)$ is not crucial for 
the formation of mounds. The evolution of the interface structure in the
mounded regime of model I is illustrated in Fig.\ref{bigfig3} which
shows the interface profiles obtained in a typical $L=200$ run starting
from a flat initial state at three different times: $t=200$ (before 
the onset of the instability), $t=4000$ (after the onset of the instability, 
in the coarsening regime), and $t=128000$ (in the final steady state).
This figure also shows the steady-state interface profile of a $L=500$ sample
with the same parameters, to illustrate that the results do not depend on the
sample size.

Very similar behavior is found for model II. Since the heights in this
atomistic model can increase
only by discrete amounts in each unit of discrete time, the increase
of $W$ at the onset of the instability is less rapid here than in the
continuum models I and IA. Nevertheless, the occurrence of the instability
for small values of $c$ shows up as a sharp upward deviation of the $W$
versus $t$ plot from the initial power-law behavior with 
$\beta \simeq 1/3$. This is illustrated by the dash-dotted line in
Fig.\ref{bigfig4}, obtained from simulations of model II with $\lambda=2.0$,
$c=0.005$. This behavior is to be contrasted with that for $\lambda=2.0$,
$c=0.015$, shown by the full line in Fig.\ref{bigfig4}, where the
nonlinear instability is absent. The difference between the surface
morphologies in the two regimes of model II is
illustrated in Fig.\ref{bigfig5}.
The kinetically rough, self-affine morphology obtained for $c=0.02$ is
clearly different from the mounded profile found for $c=0.005$. The 
time evolution of the interface in the mounded regime of this model is
illustrated in Fig.\ref{bigfig6}. The general behavior is clearly similar to
that found for models I and IA. This figure also shows a properly scaled
plot of the interface profile of a $L=500$ sample with the same parameters
at a time in the coarsening regime. It is clear from this plot that the
nature of the interface and the value of the selected slope do not depend
on the sample size.

The occurrence of a peak and a symmetrically placed trough in the
steady-state profiles shown in Figs \ref{bigfig3} and \ref{bigfig6} 
is a consequence of using periodic boundary conditions. The deterministic
part of the growth equation of Eq.(\ref{cld}) 
strictly conserves the average height if
periodic boundary conditions are used. So, the average height remains very
close to zero if the initial state is flat, as in most of our simulations.
The steady-state profile must have at least one peak and one trough in
order to satisfy this requirement. Also,
it is easy to show that if the slopes of the "uphill" and 
"downhill" parts of the steady-state profile are the same in magnitude 
(this is true for our models), then the two extrema must be separated by 
$\simeq L/2$. Therefore, it is clear that the steady state obtained in
simulations with periodic boundary conditions must have a peak and a
symmetrically placed trough separated by distance $\simeq L/2$.

To check whether the basic phenomenology described above depends on the
choice of the boundary condition, we have carried out test simulations
using two other boundary conditions:
``fixed'' boundary condition, in which the height variable 
to the left of $i=1$, and to the right of $i=L$ are pinned to zero at 
all times; and `` zero flux'' boundary condition with 
vanishing first and third derivatives of the height at the two ends of the 
sample. For these boundary conditions, the deterministic part of the 
growth equation does not strictly
conserve the average height. As a result, the symmetry between the
mound and the trough, found in the long-time steady state obtained for
periodic boundary condition, is not present if one of the
other boundary conditions is used. In particular, it is possible to
stabilize a single mound or a single trough in the steady state for the
other boundary conditions. Since the heights at the two ends must be the 
same for fixed boundary condition, the two  
extrema in a configuration with one mound and one trough must be separated 
by $\simeq L/2$, as shown in Fig.\ref{bigfig7}. 
The two extrema would not be separated by $\simeq L/2$ for 
the zero-flux boundary condition.
These effects of boundary conditions are illustrated in
Fig.\ref{bigfig7} which shows profiles in the mounded regime obtained for the
three different boundary conditions mentioned above.
It is clear from the results shown in this figure that 
the basic phenomenology, i.e. the
formation and coarsening of mounds and slope selection, is not affected
by the choice of boundary conditions. In particular, the values of 
the selected slope and the heights of the pillars at the top of a mound and the
bottom of a trough remain unchanged when boundary conditions 
other than periodic are used.

The selection of a ``magic slope'' during the coarsening process is 
clearly seen in the plots of Fig.\ref{bigfig3} and Fig.\ref{bigfig6}. 
More quantitatively, 
the probability distribution of the magnitude of the 
nearest-neighbor height differences 
$s_i \equiv |h_{i+1}-h_i|$ is found to exhibit a pronounced peak at 
the selected value of the slope, and the position of this peak does 
not change during the coarsening process. Fig.\ref{bigfig8} shows a 
comparison of the distribution of the magnitude of the 
nearest-neighbor height difference 
for model I in the mounded and kinetically rough phases. A bimodal 
distribution is seen for the mounded phase, the two peaks corresponding 
to the values of the selected slope and the height of the pillars at 
the top and bottom of the pyramids. The kinetically rough phase, on 
the other hand, exhibits a distribution peaked at zero. 
Fig.\ref{bigfig9} shows the values of the selected slope 
at different times in the coarsening regime of model I. The 
constancy of the slope is clearly seen in this plot. 
All these features remain true 
for the discrete model. Plots of the distribution $ P(s)$ at two 
different times in the coarsening regime of model II are shown in 
Fig.\ref{bigfig10}. The peak position shows a small shift in the positive
direction as $t$ is increased, but this shift is small compared to the
width of the distributions, indicating  near constancy of the selected slope. 
The value of the selected slope depends on the parameters $\lambda$ and $c$. 
This is discussed in the next section.

\section{Dynamical Phase Transition}
\label{pt}

The instability that leads to mound formation in our models is a nonlinear 
one, so that the perfectly flat state of the interface is a locally stable
steady-state solution of the zero-noise growth equation for all parameter 
values. When the instability is absent (e.g. for large values of the control
parameter $c$), this ``fixed-point'' solution of the noise-free equation
is transformed to the kinetically rough steady state in the presence of
noise. The mounded steady state obtained for small values of $c$ must 
correspond to a different fixed point of the zero-noise growth equation. Such
nontrivial fixed-point solutions may be obtained from the following simple
calculation.

The profile near the top ($i=i_0$) of a triangular 
mound may be approximated as 
$h_{i_0}=x_0+x_1,\,h_{i_0 + j} = x_0 -(|j|-1)x_2$, where $x_1$ is the 
height of the pillar at the top of the mound and $x_2$ is the selected 
slope. This profile would not change under the dynamics 
of Eq.(\ref{cld}) with no noise if the following conditions are 
satisfied:
\begin{eqnarray}
&&{\tilde{\nabla}}^2 h_{i_0} - \lambda (1-e^{-c|\tilde{\nabla} 
h_{i_0}|^2})/c \nonumber \\
&=& {\tilde{\nabla}}^2 h_{i_0\pm 1} - \lambda (1-e^{-c|\tilde{\nabla} 
h_{i_0\pm 1}|^2})/c \nonumber \\
&=& {\tilde{\nabla}}^2 h_{i_0\pm 2} - \lambda (1-e^{-c|\tilde{\nabla} 
h_{i_0\pm 2}|^2})/c. \label{cond}
\end{eqnarray}
These conditions lead to the following pair of non-linear 
equations for the variables $x_1$ and $x_2$ used to parametrize the profile
near the top of a mound:
\begin{eqnarray}
2x_1-\lambda[1-e^{-cx_2^2}]/c &=& 0, \nonumber \\
3x_1 - x_2 - \lambda [1-e^{-c(x_1+x_2)^2/4}]/c &=& 0.
\label{stability}
\end{eqnarray}
These equations admit a non-trivial solution for sufficiently small
$c$, and the resulting values of $x_1$ and $x_2$ are found to be quite
close to the results obtained from numerical integration. A similar analysis
for the profile near the bottom of a trough (this amounts to replacing
$x_2$ by $-x_2$ in Eq.(\ref{stability})) yields slightly different values 
for $x_1$ and $x_2$. The full stable profile (a fixed point of the dynamics 
without noise) with one peak and one trough may be obtained numerically by 
calculating the values of $\{h_i\}$ for which $g_i$, 
the term multiplying $\Delta t$ in the right-hand side of Eq.(\ref{cld}), is
zero for all $i$.
The fixed-point values of $\{h_i\}$ satisfy the following equations:
\begin{equation}
g_i = {\tilde{\nabla}}^2 [-{\tilde{\nabla}}^2 h_{i} + 
\lambda (1-e^{-c|\tilde{\nabla} 
h_{i}|^2})/c] = 0\,\, \hbox{for\,\,all}\,\,i.\label{fp}
\end{equation}
A numerical solution of these coupled nonlinear equations
shows that the small mismatch between the values of $x_2$ 
near the top and the bottom is accommodated by creating a few ripples near 
the top. The numerically obtained fixed-point profile for a $L=500$ system 
with $\lambda=4.0$, $c=0.02$ is shown in Fig.\ref{bigfig11}, along with a 
typical steady-state profile for the same system. The two profiles are found 
to be nearly identical, indicating that the mounded steady state in the
presence of noise corresponds to this fixed-point solution of the noiseless
discretized growth equation.

Fixed-point solutions of the continuum equation, Eq.(\ref{lds1}), with 
$\nu=1$ and $|\nabla h|^2$ replaced by $f(|\nabla h|^2)$ where $f(x)$ has
the form shown in Eq.(\ref{pvform})
may also be obtained by a semi-analytical approach following 
Racz {\it et al.}~\cite{racz}. We consider stationary 
solutions of the continuum 
equation that satisfy the following first-order differential equation
with appropriate boundary conditions:
\begin{equation}
- \frac{ds}{dx} + \lambda \frac{s^2}{1+c s^2} = A, \label{cont}
\end{equation}
where $s(x)=dh(x)/dx$ is the local slope of the interface and $A$ is a 
constant that must be positive in order to obtain a solution 
that resembles a triangular mound. 
At large distances from the peak of the mound, the slope $s$ would be
constant, so that $ds(x)/dx$ would vanish, whereas the second term 
would give a positive contribution if $\lambda$ is positive. 
At the peak of the profile, the 
second term would be  zero because $s$ is zero, but $ds(x)/dx$  
would be negative, making the left-hand side of Eq.(\ref{cont}) positive.
While a closed-form solution of this differential equation cannot be 
obtained, the value of $s(x)$ at any point $x$ may be calculated with any 
desired degree of accuracy by numerically solving a simple 
algebraic equation. The height profile is then obtained by integrating 
$s(x)$ with appropriate boundary conditions. In our calculation, we used 
the procedure of Racz {\it et. al.}~\cite{racz} to take into account periodic 
boundary conditions. In Fig.~\ref{bigfig11}, 
we have shown a typical steady-state
profile of a $L=200$ sample of model IA with $\lambda=4.0$ and $c=0.01$, and
a fixed-point solution of the corresponding continuum equation. The value of
the constant $A$ in Eq.(\ref{cont}) was chosen to yield the same 
slope as that of  the steady-state profile of the discrete model.
These results show that the steady-state properties 
for the two forms of $f(x)$ are very similar, and the continuum equation 
admits stationary solutions that are very similar to those of the discretized 
models.

The local stability of the mounded fixed point may be determined from a
calculation of the eigenvalues of the stability matrix, $M_{ij}=\partial
g_i/\partial h_j$, evaluated at the fixed point. We find that the
largest eigenvalue of this matrix (disregarding the trivial zero eigenvalue
associated with an uniform displacement of the interface, $h_i \to 
h_i + \delta$ for all $i$) crosses zero at $c=c_1(\lambda)$ (see
Fig.(\ref{bigfig12})), signaling an instability of the mounded
profile. The structure of Eq.(\ref{cld}) implies that $c_1(\lambda)
\propto \lambda^2$. Thus, for $0<c<c_1(\lambda)$, the dynamics of
Eq.(\ref{cld}) without noise admits two locally stable invariant
profiles: a trivial, flat profile with $h_i$ the same for all $i$, and a
non-trivial one with one mound and one trough. Depending on the initial
state, the noiseless dynamics takes the system to one of these two fixed
points. For example, an initial state with one pillar on a flat
background is driven by the noiseless dynamics to the flat fixed point if the
height of the pillar is smaller than a critical value, and to the
mounded one otherwise. 

The ``relevant'' perturbation that makes the mounded fixed point 
unstable at $c=c_1$ is a 
uniform vertical relative displacement of 
the segment of the interface between the  
peak and the trough of the fixed-point profile. This can be seen by numerically 
evaluating the right eigenvector corresponding to the eigenvalue of the 
stability matrix that crosses zero at $c=c_1$.
This is demonstrated in the inset of Fig.\ref{bigfig12}. Also, examination
of the time evolution of the mounded structure for values of $c$ slightly 
higher than $c_1$ shows that the instability of the structure first 
appears at the bottom of the trough. Taking cue  from these observations, 
the value $c_1$  can be obtained from a simple calculation. 
We consider the profile near the bottom of a trough at $i=i_0$. As discussed
above, the profile near $i_0$ may be parametrized as $h_{i_0} = x_0 + x_1$,
$h_{i_0+j}= x_0+(|j|-1) x_2$, and the values of $x_1$ and $x_2$ may be 
obtained by solving a pair of nonlinear equations, Eq.(\ref{stability}) with
$x_2$ replaced by $-x_2$. We now consider a perturbation of this profile,
in which the heights on one side of $i_0$ are all increased by a small
amount $\delta$ (i.e. $h_{i_0+j} = x_0+(j-1)x_2+\delta$, $h_{i_0-j} = x_0
+(j-1)x_2$ with $j> 0$), and use Eq.(\ref{cld}) to calculate how $\delta$
changes with time, assuming its value to be small. The requirement that
$\delta$ must decrease with time for the fixed-point structure to be
locally stable leads to the following equation for the value of $c$ at 
which the structure becomes unstable:
\begin{equation}
\frac{\lambda}{2} (x_1-x_2) e^{-c (x_1-x_2)^2/4} = 1, \label{asym}
\end{equation} 
By substituting the numerically obtained values of $x_1$ and $x_2$ in
this equation, the critical value, $c_1(\lambda)$, of the parameter $c$
is obtained as a function of $\lambda$.
The values obtained this way are in good agreement with those obtained 
from our full numerical calculation of the eigenvalues of the stability matrix. 
The ``spinodal'' lines (i.e. the lines in the $c-\lambda$ plane beyond 
which the mounded fixed point is unstable) for models I and IA are
shown in  Fig.\ref{bigfig13}. Both these lines have the expected form,
$c_1(\lambda) \propto \lambda^2$. It would be interesting to carry out a
similar stability analysis for the mounded stationary profile 
(see Fig.\ref{bigfig11}) of the continuum equation corresponding to model 
IA. Such a calculation would have to be performed {\it without discretizing
space} if we want to address the question of whether the behavior of the
truly continuum equation is similar to that of the discretized versions 
considered here. We have not succeeded in  carrying out such a calculation:
since the mounded 
stationary profiles for the continuum equation are obtained from
a numerical calculation, it would be extremely difficult, if not
impossible, to carry out
a linear stability analysis for such stationary solutions 
without discretizing space.

In the presence of the noise, the perfectly flat fixed point transforms
to the kinetically rough steady state, and the non-trivial fixed point
evolves to the mounded steady state shown in Fig.\ref{bigfig11}. A
dynamical phase transition at $c=c_2(\lambda) < c_1(\lambda)$ separates
these two kinds of steady states. To calculate $c_2(\lambda)$, we start
a system at the mounded fixed point and follow its evolution according to
Eq.(\ref{cld}) for a long time (typically $t=10^4$) to check whether it
reaches a kinetically rough steady state. By repeating this
procedure many times, the probability, $P_1(\lambda,c)$, of
a transition to a kinetically rough state is obtained. For fixed $\lambda$,
$P_1$ increases rapidly from 0 to 1 as $c$ is increased above a
critical value. Typical results for $P_1$ as a function of $c$ for
model I with 
$\lambda=4.0$ are shown in the inset of Fig.\ref{bigfig13}. The value of
$c$ at which $P_1=0.5$ provides an estimate of $c_2$. Another estimate
is obtained from a similar calculation of $P_2(\lambda,c)$, the
probability that a flat initial state evolves to a mounded steady state.
As expected, $P_2$ increases sharply from 0 to 1 as
$c$ is decreased (see inset of Fig.\ref{bigfig13}), and the value of $c$ at
which this probability is 0.5 is slightly lower than the value at which
$P_1=0.5$. This difference reflects finite-time hysteresis effects. The
value of $c_2$ is taken to be the average of these two estimates, and
the difference between the two estimates provides a measure of the
uncertainty in the determination of $c_2$. The phase boundary obtained
this way is shown in Fig.\ref{bigfig13}, along with the results for
$c_2(\lambda)$ obtained for the discrete model II from a similar analysis.

The general behavior found for all the models as the parameters $\lambda$
and $c$ are varied is qualitatively very similar to that in equilibrium
first order phase transitions of two- and three-dimensional systems as
the temperature and other parameters, such as the magnetic field in
spin systems, are varied. To take a standard example of an equilibrium first
order transition, we consider a system with a scalar order-parameter
field $\psi({\bf r})$, described by a Ginzburg-Landau free energy 
functional~\cite{ma} that has a cubic term:
\begin{equation}
F[\psi] = \int d{\bf r} \left[\frac{1}{2} a \psi^2({\bf r}) - \frac{1}{3} b
\psi^3({\bf r}) + \frac{1}{4} u \psi^4({\bf r})\right], \label{gl}
\end{equation}
where $b$ and $u$ are positive constants, $a=a_0(T-T_0)$ with $a_0>0$, 
and $T$ is the
temperature. Considering uniform states, $\psi({\bf r}) = m$, the free energy 
per unit volume may be written as
\begin{equation}
f(m) = \frac{1}{2} a m^2 -\frac{1}{3} bm^3 + \frac{1}{4} u m^4.
\label{mft}
\end{equation}
It is easy to show that for $T_0 < T < T_s = T_0+b^2/(4a_0u)$, the function 
$f(m)$ has two local minima, one at $m=0$, and the other at a positive value
of $m$. These two minima represent the two phases of the system. 
This system  exhibits a first order equilibrium phase
transition from the disordered phase ($m=0$) to an ordered phase with 
positive $m$ as the temperature is decreased. The transition temperature $T_c$
lies between $T_0$ and $T_s$. The temperature $T_s$ at which
the minimum corresponding to the ordered phase
disappears is called the ``spinodal'' temperature for the ordered
phase. The spinodal temperature for the disordered phase is $T_0$.

Now consider the dynamics of this system according to the following 
time-dependent Ginzburg-Landau equation~\cite{ma}:
\begin{equation}
\frac{\partial \psi({\bf r},t)}{\partial t} = -\Gamma \frac{\delta F[\psi]}
{\delta \psi({\bf r},t)} + \eta({\bf r},t), \label{tdgl}
\end{equation}
where $\Gamma$ is a kinetic coefficient and $\eta$ represents Gaussian
delta-correlated noise whose variance is related to $\Gamma$ and the 
temperature $T$ via the fluctuation-dissipation theorem~\cite{ma}. 
In the absence of noise, this equation converges to
local minima of the functional $F$. So, the noiseless dynamics exhibits two
locally stable fixed points for $T_0 < T < T_s$, corresponding to the 
two minima of $f(m)$ that represent the disordered and uniformly ordered 
states. This is analogous to the two locally stable fixed points of our 
nonequilibrium dynamical systems for $c < c_1(\lambda)$. If we identify the
flat and mounded fixed points as the ``disordered'' and ``ordered'' states,
respectively, and the control parameter $c$ to play the role of the
temperature $T$, then the noiseless dynamics of our models would look 
similar to that of Eq.(\ref{tdgl}) for $\eta=0$, with $c_1$ playing the role
of the spinodal temperature $T_s$ of the equilibrium problem.

In the presence of noise, the system described by Eq.(\ref{tdgl}) exhibits
a first-order phase transition at $T_c$ that lies between $T_s$ 
and $T_0$: the system
selects one of the phases corresponding to the two fixed points of the 
noiseless dynamics, except at $T_c$ where both phase coexist. The local 
stability of the mean-field ordered and disordered states in a small 
temperature-interval around $T_c$ is manifested in the dynamics as finite-time
hysteresis effects. The behavior we find for out nonequilibrium dynamical
models is qualitatively similar: the system selects the steady state 
corresponding to
the mounded (``ordered'') fixed point of the noiseless dynamics as the 
control parameter $c$ (analogous to the temperature $T$ of the equilibrium
system) is decreased below $c_2$ which is smaller than the ``spinodal'' 
value $c_1$. The growth models do not exhibit a ``spinodal'' point for the
kinetically rough (``disordered'') phase: the flat fixed point of the 
noiseless dynamics is locally stable for all positive values of the control
parameter $c$. If this analogy with equilibrium first order transition is
correct, then our models should show hysteresis and coexistence of kinetically
rough and mounded morphologies for values of $c$ near $c_2(\lambda)$. As
mentioned above, we do find hysteresis (see 
inset of Fig.\ref{bigfig13}) in finite-time
simulations with values of $c$ near $c_2$. Evidence for two-phase 
coexistence is presented in Fig.\ref{bigfig14}, where a snapshot of the 
interface profile for a $L=500$ sample of model I with $\lambda=4.0$, $c=0.42$
is shown. This value of $c$ is very close to the critical value $c_2$ for 
$\lambda=4.0$ (see inset of Fig.\ref{bigfig13}). This plot clearly 
illustrates the simultaneous presence of mounded and rough morphologies in 
the interface profile.

The results described above suggest that our growth models exhibit a 
{\it first-order dynamical phase transition} at $c=c_2(\lambda)$. To make this
conclusion more concrete, we need to define an {\it order parameter}, 
analogous to the quantity $m$ in the equilibrium problem discussed above,
that is zero in the kinetically rough phase, and jumps to a non-zero value
as the system undergoes a transition to the mounded phase at $c=c_2$. The
identification of such an order parameter would also be useful for
distinguishing between these two different kinds of growth in experiments --
as mentioned in the Introduction, it is difficult to experimentally
differentiate between kinetic roughening and mound formation with coarsening
from measurements of the usual bulk properties of the interface.
A clear distinction between the two morphologies may be obtained from 
measurements of the average number of extrema of the height 
profile~\cite{tor}. The steady-state profile in the mound-formation 
regime exhibits two extrema for {\it all} values of the system size $L$. 
In contrast, the number of extrema in the steady state in the kinetic 
roughening regime increases with $L$ as a power law~\cite{tor} -- we find 
that for values of $c$ for which the system is kinetically rough, e.g. 
for $\lambda = 4.0$, $c=0.05$ for model I, 
the average number of extrema in the steady 
state is proportional to $L^\delta$ with $\delta \simeq 0.83$. This 
observation allows us to define an ``order parameter'' that is zero in the 
large-$c$, kinetic roughening regime and finite in the small-$c$, 
mound-formation regime. Let $\sigma_i$ be an Ising-like variable, equal to 
the sign of the slope of the interface at site $i$. An extremum in the 
height profile then corresponds to a ``domain wall'' in the configuration 
of the $\{\sigma_i\}$ variables. Since there are two domain walls separated 
by $\sim L/2$ in the steady state in the mound-formation regime, the 
quantity 
\begin{equation}
m = \frac{1}{L} |\langle \sum_{j=1}^L \sigma_j e^{2\pi i j/L}\rangle |, 
\label{op}
\end{equation}
where $\langle \ldots \rangle$ represents a time-average in the steady state,
would be finite in the $L \to \infty$ limit. On the other hand, $m$
would go to zero for large $L$ in the kinetically rough regime because 
the number of domains in the steady-state profile would increase with $L$. 
We find numerically that in the kinetically rough phase,
$m \sim L^{-\gamma}$ with 
$\gamma \simeq 0.2$. The finite-size 
scaling data for the order parameter $m$ for models I and II for both faceted 
and kinetically rough phases is shown in Fig.\ref{bigfig15}. It is seen 
that $mL$ varies linearly with the system size $L$ in the mounded phase,
whereas
$mL \sim L^{1-\gamma}$ with $\gamma \simeq 0.2$ for model I and $\gamma 
\simeq 0.15$ for model II in the
the kinetically rough phase. So, in the $L \to \infty$ limit, the order
parameter would jump from zero to a value close to unity as $c$ is decreased
below $c_2(\lambda)$. This is exactly the behavior expected at a first-order
phase transition.

The occurrence of a first-order phase transition in our 1d models with 
short-range interactions may appear surprising -- it is well-known~\cite{ma}
that 1d systems with short-range interactions do not exhibit any 
equilibrium thermodynamic transition at a non-zero temperature. The
situation is, however, different for nonequilibrium phase transitions:
In contrast to equilibrium systems, a first-order phase
transition may occur in one-dimensional nonequilibrium systems with
short-range interactions. Several such transitions have been well documented
in the literature~\cite{noneq}. So, there is no reason to {\it a priori}
rule out the occurrence of a true first-order transition in our 1d
nonequilibrium systems. As discussed above, our numerical results strongly
suggest the existence of a true phase transition. However, since all our
results are based on finite-time simulations of finite-size systems, we
can not claim to have established rigorously the occurrence of a true
phase transition in our models. The crucial question in this context is 
whether the order parameter $m$ would be nonzero in the mounded phase in 
the $L \to \infty$ limit if the time-average in  Eq.(\ref{op}) is
performed over arbitrarily long times. Since the steady-state 
profile in this phase 
has a single mound and a single trough (this is clear from our simulations), 
the only way in which $m$ can go to zero is through strong ``phase
fluctuations'' corresponding to lateral shifts of the positions of the peak
and the trough. We do not find any evidence for such strong phase
fluctuations. We have calculated the time autocorrelation function of the
phase of the order parameter for small samples over times of the order of
$10^7$ and found that it remains nearly constant at a value close to unity
over the entire range of time. So, if such phase fluctuation eventually
make the order parameter zero for all values of $c$, then this must happen
over astronomically long times. Our finite-time simulations can not, of
course, rule out this possibility.

\section{Coarsening of mounds}
\label{coars}

During the late-stage evolution of the interface, the mounds 
coarsen with time, increasing the typical size of the triangular pyramidal
structures.
The process of coarsening 
occurs by larger mounds growing larger at the expense of the smaller 
ones while always retaining their ``magic'' slope. Snapshots of the 
system in the coarsening regime are shown in Figs \ref{bigfig16} and
\ref{bigfig17} for 
model I and model II, respectively.
The constancy of the slope during the coarsening process is clearly 
seen in these figures. As discussed in the Introduction, the constancy
of the slope implies that if the typical lateral size of a mound increases
in time as a power law with exponent $n$ ($R(t) \propto t^n$), then the 
width of the interface would also increase in time as a power law with 
the same exponent ($W(t) \propto t^\beta$ with $\beta = n$). Therefore, the
value of the coarsening exponent $n$ may be obtained by measuring the width 
$W$ as a function of time in the coarsening regime. In Fig.\ref{bigfig18},
we show a plot of the width as a function of time for model I with $\lambda
= 4.0$, $c=0.02$. It is clear from the plot that the time-dependence of 
the width is well-described by a power law with $\beta = n = 0.34 \pm 0.01$.
A similar plot for the discrete model II with $\lambda=2.0$, $c=0.005$, 
shown in Fig.\ref{bigfig19}, also
shows a power-law growth of the width in the long-time regime, 
but the value of the coarsening
exponent obtained from a power-law fit to the data is $\beta = n = 0.50 
\pm 0.01$, which is clearly different for the value obtained for
model I. This is a surprising result: model II was originally 
defined~\cite{kds} with the specific purpose of obtaining an atomistic
realization of the continuum growth equation of Eq.(\ref{lds1}), and 
earlier studies~\cite{kds,us1,us2} have shown that the
dynamical scaling behavior of this model in the kinetic roughening regime 
is the same as that of model I. Also, we have found in the present study that 
the dynamical phase transition in this model  has the 
same character as that in model I. So, the difference in the values of
the coarsening exponents for these two  models is unexpected. As noted
earlier, there is some evidence suggesting that the typical slope of the 
mounds in model II increases very slowly with time (see Fig.\ref{bigfig10}).
However, this ``steepening'', if it actually occurs, is too slow to account
for the large difference between the values of the coarsening exponents
for models I and II.

In order to understand these numerical results, we first address the 
question of why the mounds coarsen with time. This problem has certain
similarities with domain growth in spin systems~\cite{bray}.
Using the Ising variables $\{\sigma_i\}$ defined in the preceding section,
each height profile can be mapped to a configuration
of Ising spins. The coarsening of mounds then corresponds to a growth of the
typical size of domains of these Ising spins. There is, however, an 
important difference between the coarsening of mounds in our models and 
the usual domain growth problem~\cite{bray} for Ising spins. Domain growth
in spin systems is the process through which the system approaches
equilibrium from an out-of-equilibrium initial state. The dynamics
of this process may be understood in terms of arguments based on
considerations of the free energy (at
finite temperatures) or energy (at zero temperature). Such arguments
do not apply to our nonequilibrium growth models. The reason for the
coarsening of mounds in our models must be sought in
the relative stability of different 
structures under the assumed dynamics and the effects of noise. 

As discussed in the preceding section, the fixed point of Eq.(\ref{cld})
with one mound and one trough is locally stable for $c < c_1(\lambda)$. Since 
structures with several mounds and troughs approach this steady-state 
structure through
the coarsening process, it is reasonable to expect that fixed points of
the noiseless equation with more than one mounds and troughs would be 
unstable. We have numerically obtained fixed points of Eq.(\ref{cld}) 
with two mounds and troughs
for different values of the sample size and the separation between 
the peaks of the two mounds. The slope of the mounds at these fixed points is
found to be the same as that in the fixed point with one mound and one trough.
We find that the stability matrix for such fixed points always has a
real, positive eigenvalue,
indicating that the structure is unstable and would 
evolve to the stable  configuration with one mound and one trough. 
The magnitude of the positive eigenvalue of the stability matrix for
two-mounded fixed points depends on the sample size, the separation between
the peaks of the mounds
and the relative heights of the mounds in a complicated way. We have not
been able to extract any systematic 
quantitative information from these dependences.
We find a qualitative trend indicating that the magnitude of the 
positive eigenvalue decreases as the separation between the peaks of the 
two mounds
is increased. Since the time scale of the development of the instability of
two-mounded structures is
given by the inverse of the positive eigenvalue, this result is consistent 
with the expectation that the time required for two mounds to coalesce
should increase with the separation between the mounds.

These results suggest that the coarsening of the mounds in model I reflects
the instability of structures with multiple mounds and troughs. If this
is true, then coarsening of mounds should be observed in this model
even when the noise term in Eq.(\ref{cld}) is absent. To check
this, we have carried out numerical studies of coarsening in the 
noiseless version of  Eq.({\ref{cld}). In these studies, the time evolution
of an initial configuration with a pillar 
of height $h_0 > h_{min}(\lambda, c)$ at the central site of an otherwise 
flat interface is followed numerically in the presence of noise until
the instability caused by the presence of the pillar is 
well developed. The profiles obtained for different realizations of the
noise used in the initial time evolution are then used as initial 
configurations for coarsening runs without noise. The dotted line 
in Fig.\ref{bigfig18} 
shows the width versus time data obtained 
from this calculation. 
The coarsening exponent in the absence of noise is found to be the same  
($n \simeq 1/3$) as that of the noisy system, indicating that the 
coarsening in this model is driven by processes associated with the 
deterministic part of the growth equation.

We have examined the details of the process by which two mounds coalesce to
form a single one. The different steps in this process are illustrated in
the snapshots of interface profiles shown in Fig.\ref{bigfig17} where one
can see how the two mounds near the center come together to form a single
one as time progresses. First, the separation between the peaks of the 
mounds decreases
with time. When this distance becomes sufficiently small, the ``V''-shaped
segment that separates the peaks of the mounds ``melts'' to form a rough
region with many spikes. This region of the interface then self-organizes
to become the top part of a mound. 
Although the data shown in this figure
were obtained for model II, 
it also represents quite closely the process of coalescence
of mounds in models I and IA.
An estimate of the time scale associated with the second part of this 
process, during which the ``melted'' region of the interface transforms
into the top part of a mound, may be obtained in the following way.
The absolute value of the closest-to-zero eigenvalue of the stability 
matrix for the single-mounded fixed point provides an estimate of the 
time scale over which configurations close to the fixed point evolve
to the fixed point itself. In the inset of Fig.\ref{bigfig18}, we have
shown the dependence of the magnitude $\kappa_m$ of the closest-to-zero 
eigenvalue for $\lambda=4.0$, $c=0.02$ on the system size $L$.
The eigenvalue scales with the system size as $L^{-3}$, indicating 
that the time scale for the decay of fluctuations with length scale $L$ is
proportional to $L^3$. This is consistent with the observed value of the
coarsening exponent, $n \simeq 1/3$, which indicates that the time scale 
$\tau(x)$ for the coalescence of mounds separated by distance
$x$ is proportional to $x^{1/n} \sim x^3$. 

We have found very similar behavior
for the closest-to-zero eigenvalue of the stability matrix for the 
single-mounded fixed point of model IA in which a different form of the 
control function is used. Coarsening data for this model are shown in 
Fig.\ref{bigfig18}. In this model, there is a long time interval between
the onset of the instability and the beginning of power-law coarsening.
During this time interval, the interface segments 
near the tall pillars formed at the
instability organize themselves into triangular mounds. This process produces
a plateau in the width versus time plot. Eventually, however, power-law
coarsening with $n\simeq 1/3$ is recovered, as shown by the dashed line
in Fig.\ref{bigfig18}.
Since the onset of power-law coarsening in this model occurs at very late
times, we could not get coarsening data for this model over a very
wide time interval. Consequently, the calculated value of the coarsening
exponent for this model is less accurate. However, our results show quite
clearly that the ``universal'' features of the coarsening dynamics of
models I and IA are the same.

Going back to the discrete model II, we first examined its coarsening
dynamics in the absence of noise. The noiseless limit of this model is
not well-defined in the sense that there is no explicit noise term that
can be turned off. The stochasticity in this model arises from
two sources:  first, the randomness 
associated with the selection of the deposition site $i$ (the quantity
$K_i(\{h_j\})$ defined in Eq.(\ref{model2}) is calculated at this site and
at its nearest-neighbor sites);
and second, the randomness in the selection of one of the two neighbors 
of site $i$ in case of a tie in their values of $K_i$. 
In order to make the dynamics deterministic, we 
employ a parallel update scheme in which all the lattice sites, 
$i = 1, \cdots L$, are updated simultaneously instead of sequentially. This
eliminates the stochasticity arising from the choice of the sequence in
which the sites are updated.
The randomness associated with the selection of a neighbor in case of a tie
is eliminated by choosing the right neighbor if the serial number of the
occurrence of a tie, measured from the beginning of the simulation, is even,
and the left neighbor if the serial number is odd. 
With these modifications of the update
rules, the system evolves in a  perfectly deterministic way. To study
coarsening in this deterministic version of model II, we prepare an initial
structure with two identical mounds separated by distance $x_0$. The slope
of these mounds is chosen to be equal to the 
``selected'' value found in simulations of the original model. We then
study the time evolution of this structure  
according to the parallel dynamics defined above, monitoring how the
distance $x$ between the peaks of the mounds changes with time $t$. 
We find that the the value of $x$ increases initially, in order to 
accommodate a slightly higher value of the selected slope for the parallel 
dynamics. After reaching a maximum value that is $\sim 20\%$ higher than $x_0$,
the distance $x$ decreases with time, indicating that the
noiseless dynamics leads to coarsening.
Eventually, the two mounds coalesce into a single
one and the system remains in the state with one mound and one trough 
at later times. 
The slope of the mounds remains constant during the coarsening process. 
Assuming power-law coarsening with exponent $n$, the 
separation $x$ at time $t$ is expected to have the form
\begin{equation}
x(t) = (x_0^{1/n} - Ct)^n, \label{xoft}
\end{equation}
where $C$ is a constant. This form implies that the time $\tau(x_0)$
required for
the coalescence of two mounds separated by distance $x_0$ is proportional
to $x_0^{1/n}$. In Fig.\ref{bigfig20}, we have shown the time dependence
of $x(t)$  for three different initial separations, 
$x_0 = 80, 90$ and 100,  and fits of the  data to the form of Eq.(\ref{xoft}),
yielding the result 
$1/n = 2.9 \pm 0.1$. Only the data for $x$ at times larger than the time 
at which it returns to $x_0$ after the initial increase
are shown in the figure.
From these observations, we conclude that the 
coarsening exponent in the zero-noise limit of model II is the same ($n=1/3$)
as that found for the two versions of the continuum model. The observation
that the coarsening exponent for the noisy version of model II is
different from 1/3 then indicates that the effect of noise in the discrete
model is {\it qualitatively different} from that in the continuum models.
We discuss below a possible explanation of this behavior.

The fact that noiseless versions of all three models exhibit
the same value of the coarsening exponent ($n =1/3$) suggests that 
the coarsening is driven by an effective attractive interaction between the 
peaks of neighboring mounds. The observed value of $n$ suggests~\cite{rb} 
that the
this attractive interaction is proportional to $1/x^2$ where $x$ is the 
separation between the mound tips. This interaction would lead to the 
observed result, $\tau(x) \propto x^3$, in the noiseless limit if the
rate of change of $x$ with $t$ is assumed to be proportional to the 
attractive force (``overdamped limit''). The presence of noise in the 
original growth model leads to a noise term in the equation of motion of 
the variable $x$, but the nature of this noise term is not clear. Since
the observed coarsening dynamics in the noisy model II ($n \simeq 0.5$)
suggests a similarity with random walks, we propose that the effective
dynamics of the variable $x$ is governed by the kinetic equation
\begin{equation}
\frac{d x}{d t} = - C/x^2 + \eta(t), \label{bw} 
\end{equation}
where $\eta$ is a Gaussian, delta-correlated noise with zero mean and 
variance equal to $2D$. In this phenomenological description, 
the coarsening of a two-mounded structure in model II is mapped to 
a Brownian walk of a particle in an attractive 
potential field with an absorbing wall at the origin, such that the 
particle cannot escape once it arrives at the origin. The  
absorption of a particle at the origin corresponds to the coalescence of 
two mounds in the original height picture. Thus, in this reduced model, the 
quantity of interest is the typical first passage time $\tau$ (i.e. the 
time taken by a particle to reach the origin) as a function of $x_0$, the
initial distance of the particle from the origin. In the noiseless limit
($D = 0$), $\tau$ is equal to $x_0^3/(3C)$, and in the purely
Brownian walk limit ($C=0$), the typical value of $\tau$ should be
of the order of $x_0^2/D$. Therefore, for sufficiently large values of 
$x_0$, random-walk behavior characterized by $n=1/2$ is expected.
However, for relatively small values of $x_0$, the behavior should be 
dominated by the attractive interaction. Therefore, we expect 
that the dynamics
described by Eq.(\ref{bw}) with nonzero $C$ and $D$ would exhibit
a crossover from a noise dominated regime to an interaction 
dominated regime as the value of $x_0$ is decreased. This crossover is
expected to occur near $x_0 = x_c \sim C/D$, for which the values
of $\tau$ obtained from the two individual terms in Eq.(\ref{bw}) become
comparable. We, therefore,
propose a scaling form for the dependence of $\tau$ on $x_0$:
\begin{equation} 
\tau(x_0) = \frac{x_0^3}{3 C} F\left(\frac{ D x_0}{ C}\right)
\end{equation} 
where the scaling function $F(z)$ has the following asymptotic 
dependence on its argument $z$: 
$F(z) = 1$ for $ z \ll 1$ and $F(z) \propto 1/z$ for $z \gg 1$. Our numerical
study of the reduced model of Eq.(\ref{bw}) confirms the validity of this
scaling ansatz. Note that for any nonzero value of $D$, $\tau(x_0)$ would
be proportional to $x_0^2$, implying $n=1/2$, for sufficiently large values
of $x_0$.

To test the validity of this reduced description of the coarsening dynamics
of the original model II with stochastic sequential updates, we have simulated
the evolution of a two-mounded structure in this model. The two-mounded 
structure used in these simulations is identical to that used in the study
of coarsening in the model with parallel updates. In these simulations also,
the average value of $x$ exhibits a small initial increase followed by a
steady decrease. The initial growth of $\langle x^2(t) \rangle - \langle
x(t) \rangle^2$ with time is found to be linear, indicating the presence of
a random additive noise in the effective equation of motion of $x$. 
Fig.\ref{bigfig21} shows the time dependence of $\langle x(t) \rangle$
obtained from simulations of $L=500$
samples of model II with $\lambda=2.0$, $c=0.005$, and
$x_0 = 100$. In this plot, the origin of time has been shifted to the point
where $\langle x \rangle$ returns to the initial value $x_0 = 100$ 
after the small initial increase, and $10^3$ units of simulation 
``time'' (number of deposited layers) is taken to be
the unit of $t$. The number of independent runs used in the calculation of
averages is 800. The observed dependence of $\langle x(t) \rangle$ on $t$
can be described reasonably well by the reduced equation of Eq.(\ref{bw})
for appropriate choice of the values of the 
parameters $C$ and $D$. As shown in 
Fig.\ref{bigfig21}, the  $\langle x(t) \rangle$ calculated numerically from
Eq.(\ref{bw}) with $C=285.0$ and $2D=0.3$ provides a good fit to the data
obtained from simulations of the growth model. Due to the limited range of
the simulation data, the values of $C$ and $D$ can not be determined very 
accurately from such fits: values of $C$ in the 
range $250 - 300$ and values of $D$ in 
the range $0.1 - 0.5$ (larger values of $C$ have to be combined 
with smaller values of $D$)
provide reasonable descriptions of the simulation data. For such values
of $C$ and $D$, and $x_0 \approx L$ where $L=10^3$ is the sample size used
in the calculation of the coarsening exponent, $Dx_0/C$ is of order
unity, indicating that
the effects of the noise term in  Eq.(\ref{bw}) should be observed in the
simulation data. We, therefore, conclude that  
the presence of an additive random noise term
in the effective equation of motion for the separation between the peaks
of neighboring mounds in model II is a plausible explanation for 
the observed value of the coarsening exponent, $n =1/2$.

In view of this conclusion, it is interesting to enquire why the coarsening
exponent $n$ for models I and IA has the value $1/3$ characteristic of 
dynamics governed by the deterministic interaction between mound tips. We can
not provide a conclusive answer to this question. One possibility is that the
additive random noise in the original growth equations for these models
does not translate into a similar noise in the effective equation of motion
for the separation of mound tips. A second possibility is that the equation
of motion for the separation $x$ for these models also has the form of 
Eq.(\ref{bw}), but the relative strength of the noise is much smaller, 
so that the
crossover value $x_c$ is much larger than the typical sample sizes used
in our simulations. Under these circumstances, the dynamics of $x$ would
be governed by the interaction and $\tau$ would be proportional to $x_0^3$,
giving the value $1/3$ for the coarsening exponent $n$. If the second
explanation is correct, then one should observe a crossover from $n=1/3$
to $n=1/2$ in models I and IA as the sample size $L$ is increased. We do
not find any evidence for such a crossover in our simulations.

In passing, we note that the dynamics of the slope variables $\{s_i\}$ is 
strictly conserved in the models studied here if periodic or fixed boundary
conditions are used. Also, the deterministic part of the growth equations
conserves the integrated height. The ``magnetization'' of the Ising variables
$\{\sigma_i\}$ representing the signs of $\{s_i\}$ is not strictly conserved:
it is conserved only in a statistical sense.  
However, unlike other well-known examples~\cite{bray} of systems with
conserved dynamics, we obtain in model II 
a coarsening exponent that is different from the expected Lifshitz-Slyozov
value, $n=1/3$. Since the height variables in model II are integers, there
are some sites for which $h_i = h_{i+1}$. The assignment of the value of
the Ising variable $\sigma_i$ at such sites is clearly ambiguous. It may
be more appropriate to use a three-state variable, taking the values $\pm 1$
and $0$, to describe the coarsening behavior of this model. This problem
does not arise in models I and IA for which the height variables are real
numbers because the likelihood of two neighboring height variables to be
exactly the same is vanishingly small.

\section{Model I with conserved noise}
\label{conserv}

As discussed in section \ref{pt}, the properties of the mounded phase of
model I are determined to a large extent by the mounded fixed point of the
deterministic part of the equations of motion of the height variables.
The presence of noise changes the critical value of the control parameter
$c$ from $c_1$ to $c_2<c_1$, but does not affect strongly the properties of 
the mounded steady state of the system (see, for example, Fig.\ref{bigfig11}).
Therefore, we expect that the properties of the mounded phase would not
change drastically if the statistics of the noise is altered. On the other
hand, it is well-known~\cite{rev1,rev2} that the exponents that describe the
scaling behavior in the kinetically rough phase depend strongly on the 
nature of the noise. In particular, the exponents for conserved noise are
expected to be quite different from those describing the behavior for
nonconserved noise. The occurrence of the nonlinear instability that
leads to the mounded phase in our models is contingent upon the 
spontaneous formation of pillars of height $h_0 > h_{min}(\lambda, c)$, 
if the initial state of the system is completely flat. The
probability of formation of such pillars depends crucially~\cite{us2}
on the values of the roughening exponents which, in turn, are strongly
dependent on the nature of the noise. Therefore, we expect that the
nature of the noise may be very important in  determining whether
the instability leading to mound formation actually occurs 
in samples with flat  initial states.

We have investigated this issue in detail by carrying out simulations of
a version of Model I in which the noise is conserved~\cite{sgg}. The 
equations of motion for the height variables in this model have the
form of Eq.(\ref{cld}), with the noise terms $\{\eta_i(t)\}$ having the
properties
\begin{equation}
\langle \eta_i(t) \rangle = 0, \,\, \langle \eta_i(t) \eta_j(t^\prime) \rangle
= - {\tilde{\nabla}}^2 \delta_{i,j} \delta_{t,t^\prime},
\label{conseq}
\end{equation}
where $\delta_{ij} = 1$ if $i=j$ and zero otherwise. This model is expected
to exhibit kinetic roughening with exponents $\beta \simeq 1/11$, $\alpha
\simeq 1/3$, and $z \simeq 11/3$ in one dimension~\cite{sgg}. Since the
value of $\alpha$ for this model is less than unity, it
exhibits conventional dynamical scaling with the typical value of the 
nearest-neighbor height difference saturating at a constant value at long times.
The value of this constant is expected to increase~\cite{us2} as the 
strength $\lambda$ of the nonlinearity is increased. As discussed in detail
in Ref.\onlinecite{us2}, the nonlinear instability that leads to mound
formation is expected to occur in
the time evolution of such models from a perfectly flat initial state only
if the value of $\lambda$ is sufficiently large to allow the 
spontaneous, noise-induced formation of pillars of height greater than
$h_{min}(\lambda,c)$. So, if the value of $\lambda$ is sufficiently small,
then the model with conserved noise, evolving from a flat initial state, 
would not exhibit the mounding transition. On the other hand, if the
instability is induced in the model 
by starting the time evolution from a state in which there is
at least one pillar with height greater than $h_{min}$, then it is expected
to evolve to the mounded state if the value of $c$ is sufficiently
small to make the mounded state stable. So, the long-time 
steady state of the conserved model is expected to exhibit an interesting
dependence on the initial state: if $\lambda$ is sufficiently small (so that
pillars with height greater than $h_{min}$ are not spontaneously generated 
in the time evolution of the interface from a flat initial state),
and the value of $c$ sufficiently small (so that the mounded state is 
stable in the presence of noise), then the steady state would be kinetically
rough if the initial state is sufficiently smooth, and mounded if the initial
state contains pillars of height greater than $h_{min}$. This ``bistability''
does not occur for the nonconserved model I because the nearest-neighbor
height difference in this model continues to increase with time, so that
the instability always occurs at sufficiently long times~\cite{us2}.

Our simulations of the model with conserved noise show the bistable
behavior discussed above in a large region of the $c-\lambda$ plane.
We find that in this model, 
the height of a pillar on an otherwise flat interface 
increases in time if its initial value $h_0$ is larger than $h_{min} 
\simeq 20/\lambda$ (the dependence of $h_{min}$ on $c$ is weak). 
This dependence of $h_{min}$ on $\lambda$ is very similar to that~\cite{us2}
found for model I with nonconserved noise. We also find that the typical 
values of the nearest-neighbor height difference do not continue to
increase with time in this model. Consequently, if $\lambda$ is sufficiently
small, pillars with height greater than $h_{min}$ are not generated, and the
system exhibits conventional kinetic roughening with exponent values close
to the expected ones~\cite{sgg}. On the other hand, if the time evolution
of the same system is started from a state with a pillar of height greater 
than $h_{min}$, then it evolves to a mounded state very similar to the one
found in the model with nonconserved noise if the value of $c$ is
sufficiently small. The two steady states obtained for the conserved
model with $\lambda=4.0$, $c=0.01$ are shown in Fig.\ref{bigfig22}. 
The long-time state obtained in a run starting from a flat configuration 
is kinetically rough, whereas the state obtained in a run 
in which the nonlinear 
instability is initially seeded in the form of a single pillar of height 
$h_0=1000$ at the central site is mounded with a well-defined slope, as in 
model I with nonconserved noise. The difference between the two profiles,
obtained for the same parameter values for two different initial states,
is quite striking.

Since the steady state in the conserved model depends on the initial
condition, it is not possible to draw a conventional phase diagram for this
model in the $c -\lambda$ plane: the transition lines are different for 
different initial conditions. In Fig.\ref{bigfig23}, we have shown three
transition lines for this model in the $c-\lambda$ plane. The line drawn
through the circles (line 1) is obtained from simulations in which the
system is started from a flat initial state. If $\lambda$ is small, then
the steady state reached in such runs is kinetically rough for all
values of $c$. As
$\lambda$ is increased above a ``critical value'' $\lambda_c \simeq 5.3$,
pillars with height greater than $h_{min}$ are spontaneously generated
during the time evolution of the system and it exhibits a transition to
the mounded state if the value of $c$ is not very large. The circles
represent the values of $c$ for which 50\% of the runs show transitions to
the mounded state. The line through the diamonds (line 2) corresponds to
50\% probability of transition to the mounded state from an initial state 
with a pillar of height $h_0 =$ 1000 on an otherwise flat interface. The
probability of reaching a mounded steady state in such runs decreases from 
unity as the value of $c$ is increased, and falls below 50\% as line 2
is crossed from below. For large $\lambda$, lines 1 and 2 merge together.
This is expected: the probability of occurrence of a 
mounded steady state should not depend on how the pillars that initiate
the nonlinear instability are generated. The third line (the one passing
through the squares) represents 50\% probability of transition to the 
kinetically rough state from a mounded initial state (the fixed point of
the noiseless equations of motion with one mound and one trough). This line
reflects the noise-induced instability of the mounded steady state for
relatively large values of $c$. The differences between lines 2 and 3
are due to finite-time hysteresis effects similar to those discussed in 
section~\ref{pt} in the context of determining the critical value 
$c_2(\lambda)$ of the control parameter $c$ for model I with nonconserved
noise.

The interesting region in the ``phase diagram'' of Fig.\ref{bigfig23} is the
area enclosed by lines 1 and 2 and the $c=0$ line. For parameter values in
this region, the system exhibits bistable behavior, as discussed above. This
bistability is unexpected in the sense that in most studies of nonequilibrium
surface growth, it is implicitly assumed that the long-time steady state
of the system does not depend on the choice of the initial state. So, it is
important to examine whether the dependence of the steady state on the
initial condition in the conserved model reflects a very long (but finite)
transition time from one of the two apparent steady states to the other one.
We have addressed this question by carrying out long ($t$ of the order of 
$10^7$) simulations of small samples with  parameter values in the middle
of the ``bistable region'' for flat and mounded initial states. We did
not find any evidence for transitions from one steady state to the other one
in such simulations. Of course, we can not rule out the possibility that
such transitions would occur over much longer time scales.
 
\section{Summary and discussions}
\label{summ}

To summarize, we have shown from numerical simulations that a class of 1d
surface growth models exhibits mound formation and power-law coarsening of
mounds with slope selection as a result of a nonlinear instability that
is controlled by the introduction of an infinite series of terms with 
higher-order gradient nonlinearities. The models considered here are
discretized versions of a well-known continuum growth equation and an
atomistic model originally formulated for providing a discrete realization
of the growth equation. We have shown that these models exhibit a dynamical
phase transition between a kinetically rough phase and a mounded phase as
a parameter that measures the effectiveness of controlling the instability
is varied. We have defined an order parameter for this first-order
transition and used finite-size scaling to demonstrate how the sample-size
dependence of this order parameter provides a clear distinction between the
rough and mounded phases. We have also mapped out the phase boundary that 
separates the two phases in a two-dimensional parameter space. 

We would like to emphasize that the ES mechanism, commonly believed to be
responsible for mound formation in surface growth, is not present in our 
models. Our models exhibit a nonlinear instability, instead of the linear
instability used conventionally to represent the effects of ES barriers.
The mechanism of mound formation in our models is also different from a
recently discovered~\cite{pierre,ramana} one involving fast edge diffusion,
which occurs in two or higher dimensions. The slope selection found in our
models is a rare example of pattern formation from a nonlinear instability.
This is clearly different from slope selection in ES-type models in which
the mounds maintain a constant slope during coarsening 
only if the nonequilibrium surface current vanishes
at a particular value of the slope. The selected slope in such models is 
simply the slope at which the current is zero. The behavior of our models
is more complex: in these models, the surface current is zero for all values
of constant slope, and the selected value of the slope is obtained from
a nonlinear mechanism of pattern selection.

Our studies bring out two other unexpected results. We find that the coarsening
behavior of an atomistic model (model II) specifically designed to provide
a discrete realization of the growth equation that leads to model I is 
different from that of model I: the exponents that describe the power-law
coarsening are different in the two models. We show that this difference may 
arise from a difference in the nature of the effective noise that
enters the equation of motion for the separation between neighboring 
mounds in the  two cases. The second surprising result is
that the numerically obtained 
long-time behavior of model I with conserved noise in a region
of parameter space depends crucially on the initial conditions: the
system reaches a mounded or kinetically rough steady state depending on 
whether or not the initial state is sufficiently rough. To our knowledge,
this is the first example of ``nonergodic'' behavior in nonequilibrium
surface growth.

The behavior found in our 1d models may be relevant to experimental studies
of the roughening of steps on a vicinal surface. As noted earlier, the form
of the control function $f(x)$ used in model IA is physically reasonable.
However, since very little is known about the values of the model parameters
appropriate for experimentally studied systems, we are unable to determine
whether the mechanism of mound formation found in our study would be
operative under experimentally realizable conditions. The nonlinear
instability found in our 1d models is also present~\cite{kund}
in the experimentally 
interesting two-dimensional version of the growth equation of  Eq.(\ref{lds1}).
However, it is not clear whether this instability, when controlled in a
manner similar to that in our 1d models, would lead to the formation of
mounds in two dimensions. This question is currently under investigation.
Since the growth equation of Eq.(\ref{lds1}) exhibits conventional dynamic
scaling in the kinetic roughening regime in two dimensions, the nonlinear
instability would not occur in runs with flat initial states if the value
of $\lambda$ is small. Therefore, the behavior in two dimensions is
expected to be similar to that of our 1d model I 
with conserved noise: the nature
of the long-time steady state may depend crucially on initial conditions
in a region of parameter space. Such nonergodic behavior, if found in two
dimensions, would have interesting implications for the growth of films on
patterned substrates.

All the results described in this paper have been obtained from numerical
studies of models that are discrete in both space and time. 
It is interesting to enquire
whether the truly continuum growth equation of Eq.(\ref{lds1}) exhibits 
similar behavior. This question acquires special significance in view of
studies~\cite{us2,nbray} that have shown that discretization may
drastically change the behavior of nonlinear growth equations similar to
Eq.(\ref{lds1}). Since the interesting behavior found in our discretized
models arises from the nonlinear instability found earlier~\cite{us1,us2}, the
question that we have to address is whether a similar instability is present
in the truly continuum growth equation.
This question was addressed in some detail in Ref.\cite{us2} where it was shown
that the nonlinear instability is not an artifact of discretization of time or
the use of the simple Euler scheme for integrating the the growth equation.
In the present study, we have found additional evidence (see 
section~\ref{mound}) that supports this claim. We should also point out that
the atomistic model II, for which the question of inaccuracies arising from 
time integration does not arise, exhibits very similar behavior, suggesting 
that the behavior found in models I and IA is not an artifact of the time
discretization used in the numerical integration.

The occurrence of the nonlinear instability does depend on the way space 
is discretized (i.e. how the lattice derivatives are defined). In earlier 
work~\cite{us1,us2} as well as in the present study, 
the lattice derivatives are defined in a left-right 
symmetric way. We have found that the instability actually becomes 
stronger if the number of neighbors used in the definition of the lattice
derivatives is increased. 
This result suggests that the instability is also present in the continuum 
equation. It has been found by Putkaradze {\it et al.}~\cite{put} that the 
instability does not occur if the lattice derivatives are defined in a 
different way in which either left- or right-discretization of the nonlinear 
term is used, depending on the sign of the local slope of the interface. 
There is no reason to believe that this definition is ``better'' or 
''more physical'' than the symmetric definitions used in our work. 
The only rigorous result we are aware of for the behavior of Eq.(\ref{lds1})
is derived in Ref.\cite{put}  where
it is shown that the solutions of the noiseless equation are bounded for 
sufficiently smooth initial conditions. This result, however, does not answer 
the question of whether the instability  occurs 
in the continuum equation. As discussed in Ref.~\cite{us2}, the nonlinear  
instability of Eq.(\ref{lds2}), signalled by a rapid initial growth of the 
height (depth) of isolated pillars (grooves), may not lead to a true 
divergence of the height variables. The results reported in the present work  
would remain valid as long as high pillars or deep grooves are formed by the 
instability -- the occurrence of a true divergence is not necessary.

In the present work, we have shown (see section~\ref{pt}) that the continuum 
equation with $f(x)$ defined in Eq.(\ref{pvform}) does admit stationary 
solutions that exhibit all the relevant features of stationary solutions of 
the discretized equation. This result provides additional support to the
contention that the 
behaviors of the continuum and discretized systems are qualitatively 
similar. We should, however, mention that these stationary solutions of the
continuum equation do not pick out a selected slope of the interface: 
profiles similar to those shown in Fig.\ref{bigfig11} may be obtained for
different values of the parameter $A$ in Eq.(\ref{cont}) that determines the 
slope of the triangular mound. Slope selection in the continuum equation
may occur as a consequence of the requirement of local stability of such
stationary solutions. As mentioned in section~\ref{pt}, we 
have not attempted a linear stability analysis of such numerically obtained 
stationary solutions of the continuum equation because doing such a 
calculation without discretizing space would be extremely difficult. Further
investigation of this question would be useful.

Finally, we would like to emphasize that the discrete models studied here would 
continue to be valid models for describing nonequilibrium surface 
growth even if the behavior of the truly continuum growth equation of 
Eq.(\ref{lds1}) turns out to be different from that found here. 
These models may be looked upon as ones in which continuous (in models I and 
IA) or discrete (in model II) height variables defined on a 
discrete lattice evolve in continuous or discrete time. 
These models have all the correct symmetries and conservation laws of the 
physical problem, and they exhibit,
for different values of the control parameter $c$, both the phenomena 
of kinetic roughening and
mound formation found in experiments. There is no compelling reason to
consider the continuum equation to be more ``correct'' or ``physical'' 
than these models.
Epitaxial growth is intrinsically a discrete process at the molecular
level and a continuum description is an approximation that may not be
valid in some situations. 

From a different perspective, the  nonequilibrium first-order
phase transition found in our models is interesting, especially because it 
occurs in 1d systems with short range interactions. Such phase transitions
have been found earlier in several 1d ``particle hopping'' models~\cite{noneq}. 
It would be interesting to explore possible connections between such models
and the 1d growth models studied here.

\section{Acknowledgement}
\label{acknow}
We thank SERC, IISc for computational facilities, and S. Das Sarma and
S. S. Ghosh for useful discussions.

\newpage
\vbox{
\epsfxsize=8.0cm
\epsfysize=6.0cm
\epsffile{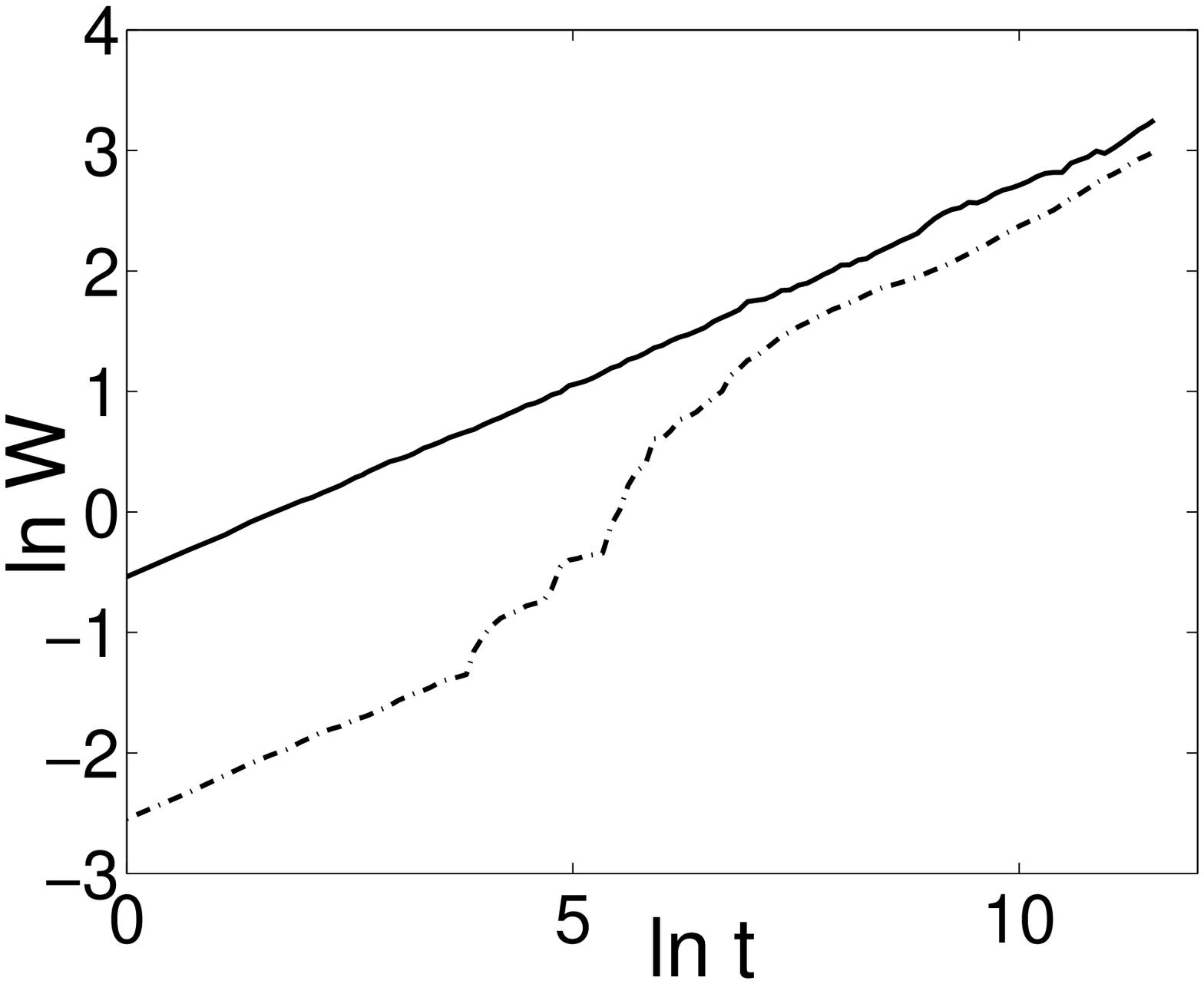}
\begin{figure}
\caption{\label{bigfig1} Double-log plots of the 
interface width $W$ as a function of time $t$ for 
model I with $\lambda=4.0$, $c=0.02$ 
(dash-dotted line), and for $\lambda=4.0$, 
$c=0.06$ (full line). These data are for $L=500$ samples, averaged 
over 40 independent runs starting from flat initial states. Plots 
have been shifted vertically for clarity.}
\end{figure}}

\vbox{
\epsfxsize=8.0cm
\epsfysize=6.0cm
\epsffile{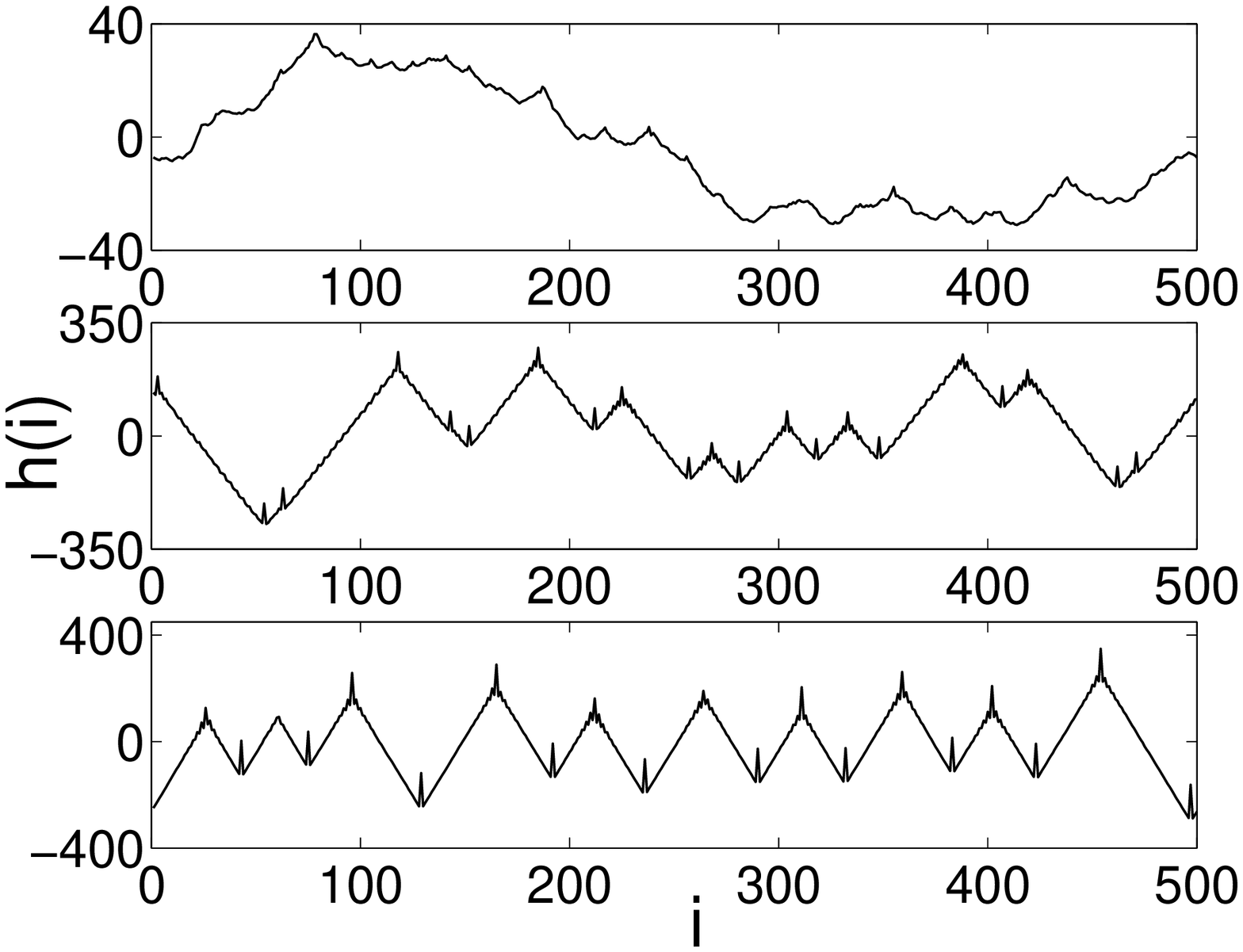}
\begin{figure}
\caption{\label{bigfig2} Configuration snapshots at 
$t$=$10^5$, for model I with $\lambda$=4.0, 
$c$=0.06 (top panel), model I with $\lambda$=4.0, $c$=0.02 
(middle panel), and model IA with $\lambda$=4.0, 
and $c$=0.01 (bottom panel).}
\end{figure}}

\vbox{
\epsfxsize=8.0cm
\epsfysize=6.0cm
\epsffile{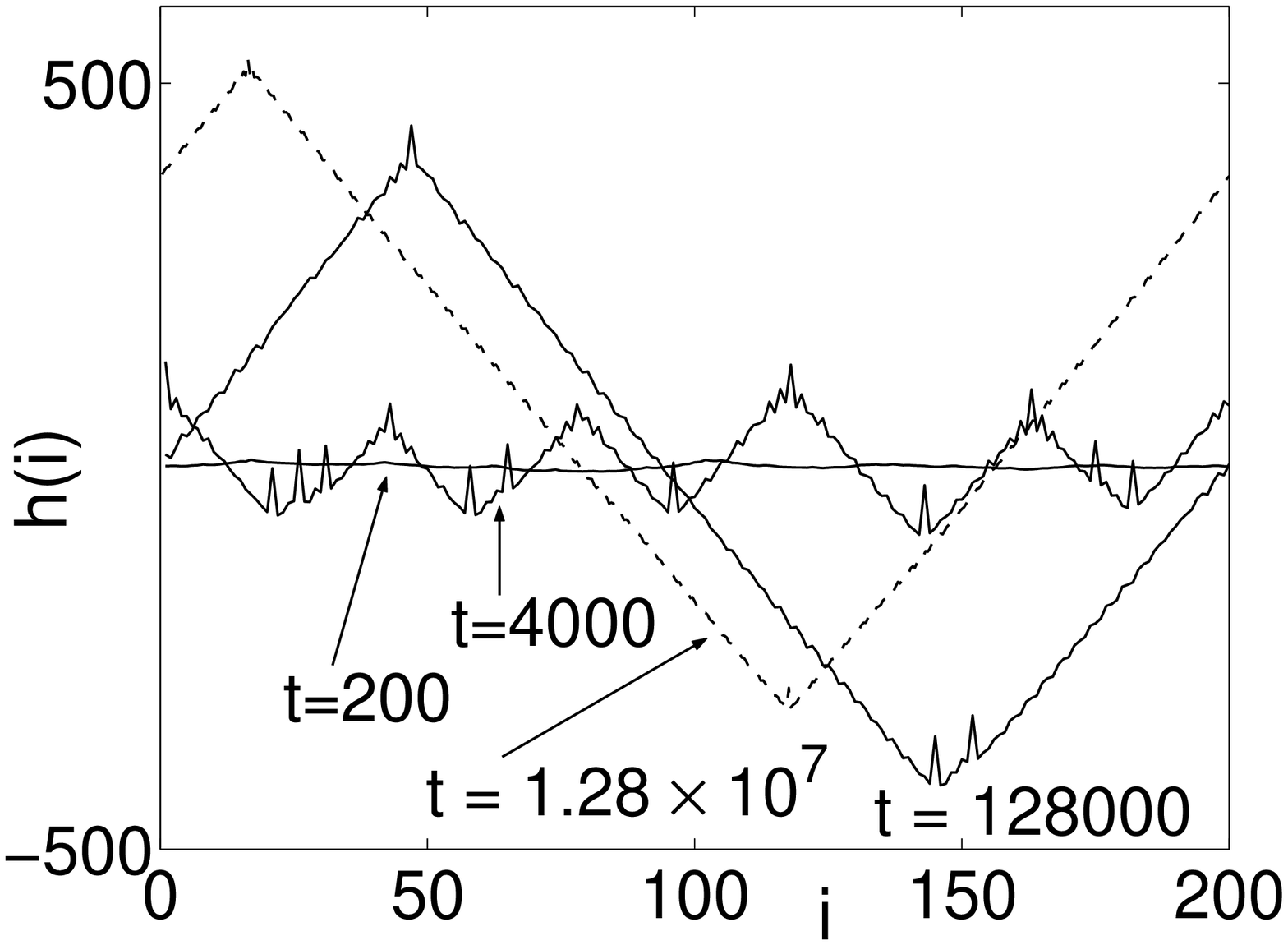}
\begin{figure}
\caption{\label{bigfig3} The interface profile 
at three different times ($t$ = 200, 4000, and 128000) 
in a run starting from a flat state for a $L$ = 200 
sample of model I with $\lambda$=4.0 and $c$=0.02. The dashed line shows
the profile for a $L=500$ sample with the same parameters 
at $t=1.28\times10^7$, with both axes
scaled by 2.5.}
\end{figure}}

\vbox{
\epsfxsize=8.0cm
\epsfysize=6.0cm
\epsffile{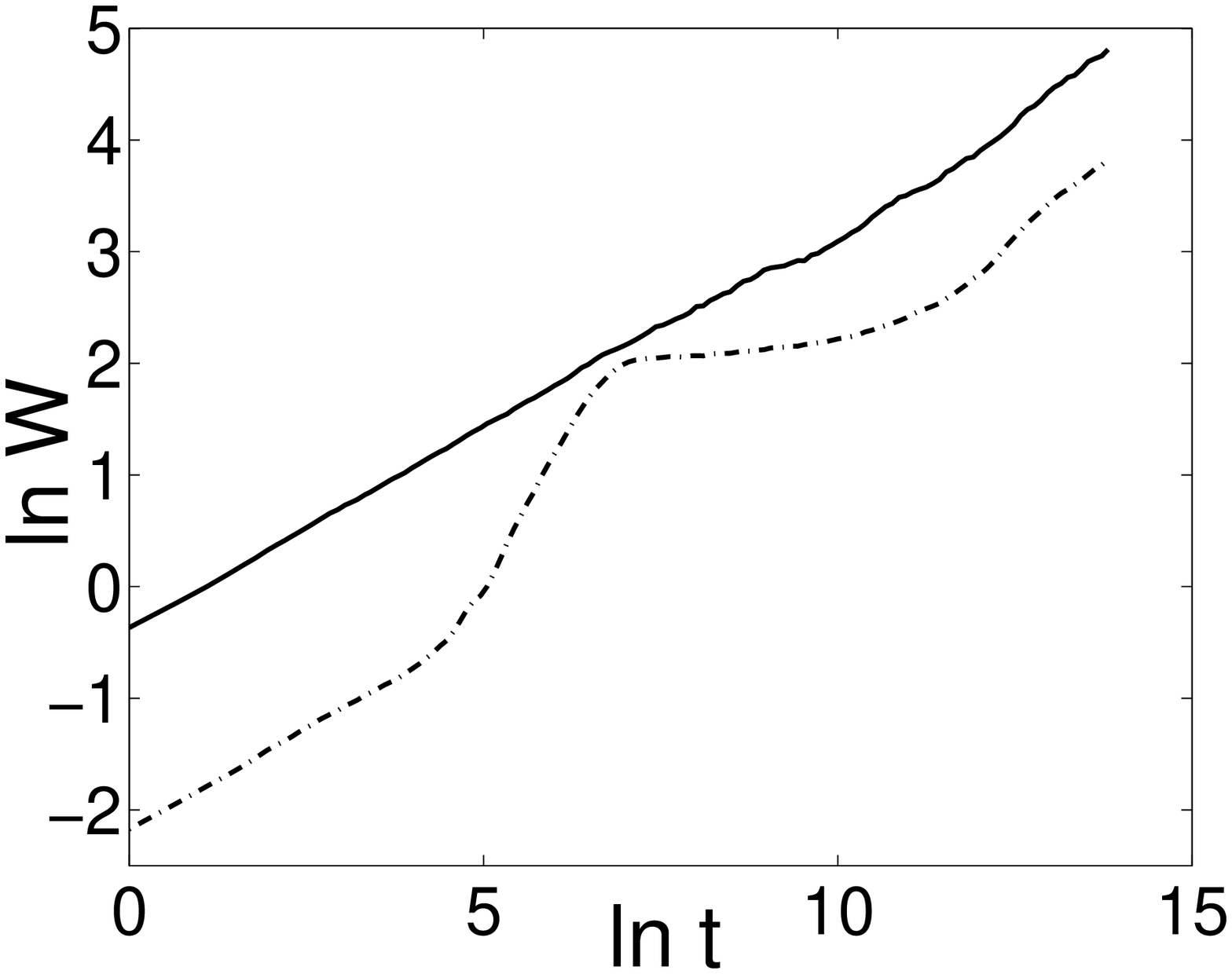}
\begin{figure}
\caption{\label{bigfig4} Double-log plots of the 
interface width $W$ as a function of time $t$ for 
model II with $\lambda$=2.0, $c$=0.005 
(dash-dotted line), and for 
$\lambda$=2.0, $c$=0.015 (full line). These data are for $L=500$ 
samples, averaged over 40 independent runs starting from flat states. Plots 
have been shifted vertically for clarity.}
\end{figure}}

\vbox{
\epsfxsize=8.0cm
\epsfysize=6.0cm
\epsffile{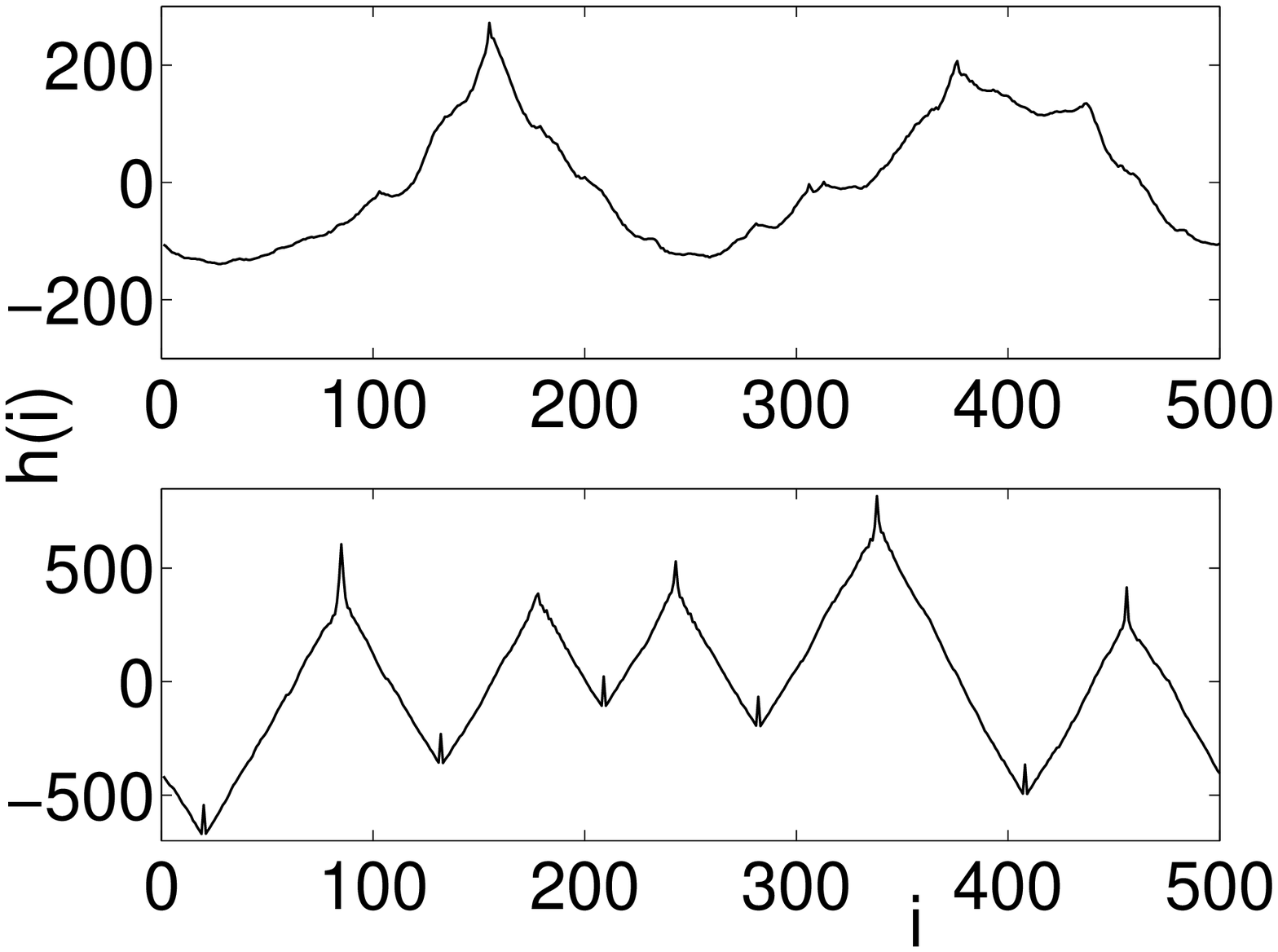}
\begin{figure}
\caption{\label{bigfig5} Configuration snapshots 
at $t$=$10^6$ for model II with $\lambda$=2.0, 
$c$=0.015 (top panel), and $\lambda$=2.0, $c$=0.005 
(bottom panel).}
\end{figure}}

\vbox{
\epsfxsize=8.0cm
\epsfysize=6.0cm
\epsffile{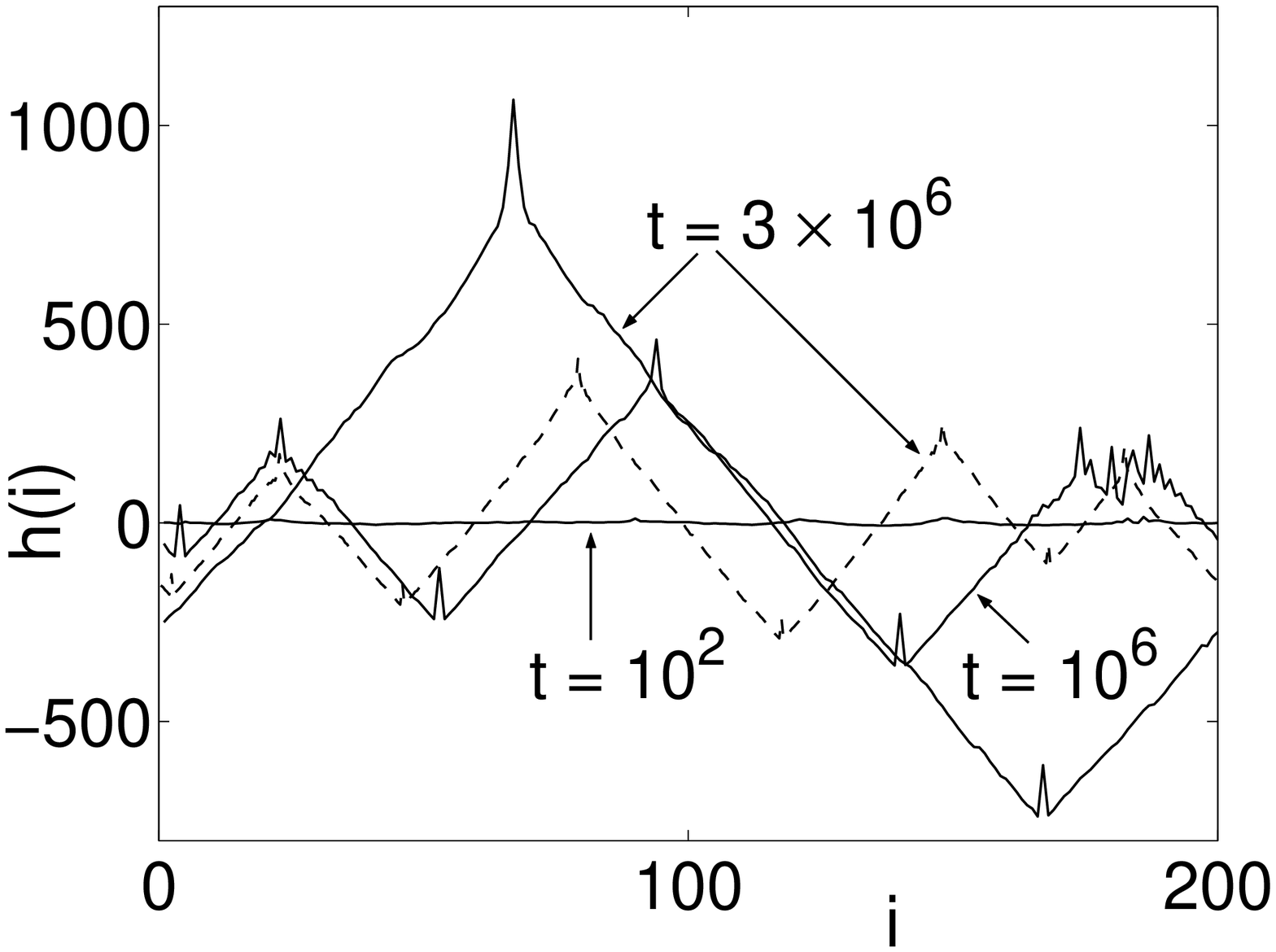}
\begin{figure}
\caption{\label{bigfig6} The interface profile at three times
($t = 10^2,\, 10^6$, and $3\times10^6$) in a run starting from a flat state
for a $L$ = 200 sample of model II with $\lambda$=2.0 and $c$=0.005.
The profile of a $L=500$ sample with the same parameters 
at $t=3\times 10^6$, with both axes scaled
by 2.5, is shown by the dashed line.}
\end{figure}}

\vbox{
\epsfxsize=8.0cm
\epsfysize=6.0cm
\epsffile{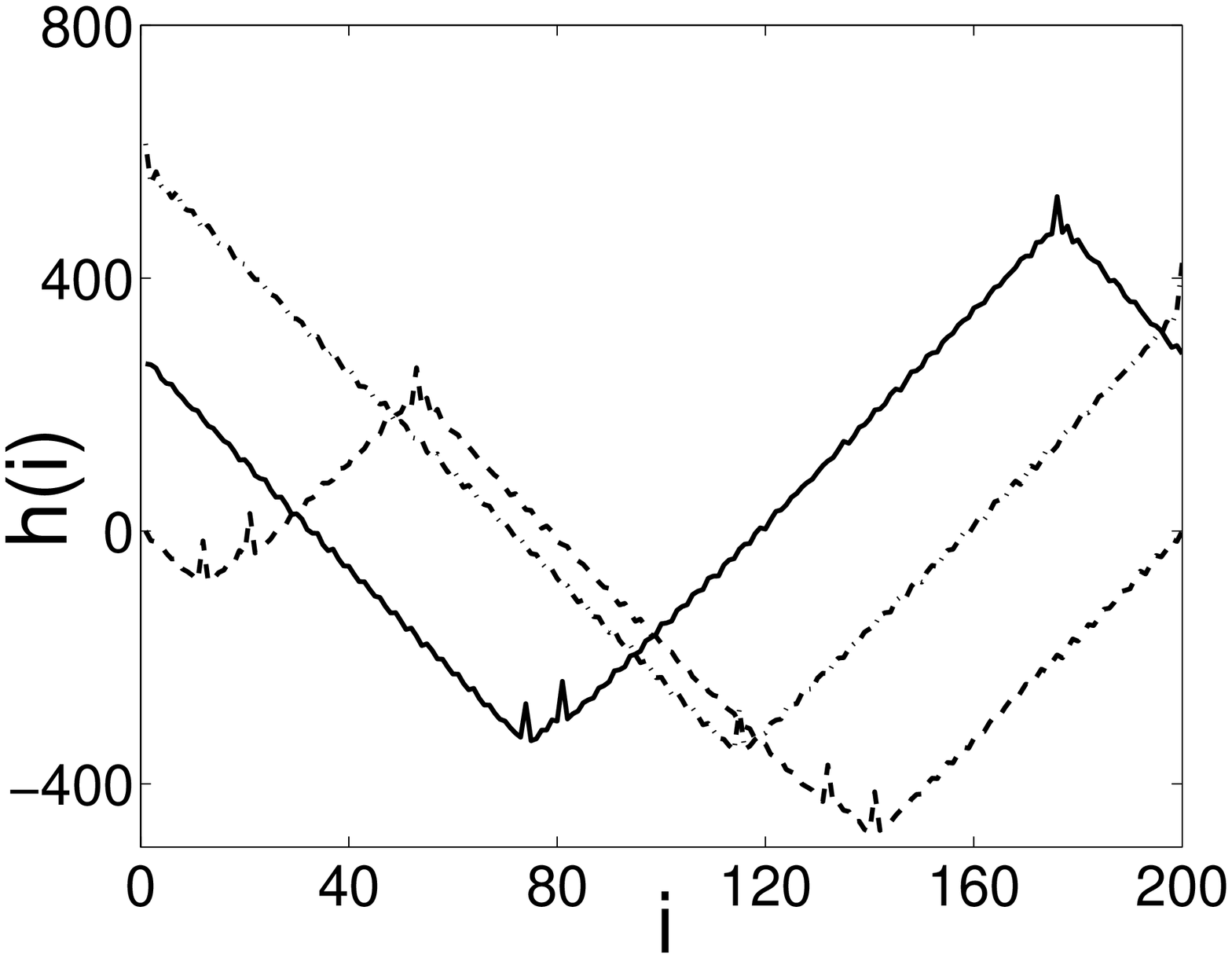}
\begin{figure}
\caption{\label{bigfig7} Height profiles for model I ($\lambda=4.0$, 
$c=0.02$) at time 
$t=1.28\times 10^7$ for 
periodic boundary conditions (full line), fixed boundary 
conditions (dashed line), and zero-flux boundary 
conditions (dash-dotted line).}
\end{figure}}

\vbox{
\epsfxsize=8.0cm
\epsfysize=6.0cm
\epsffile{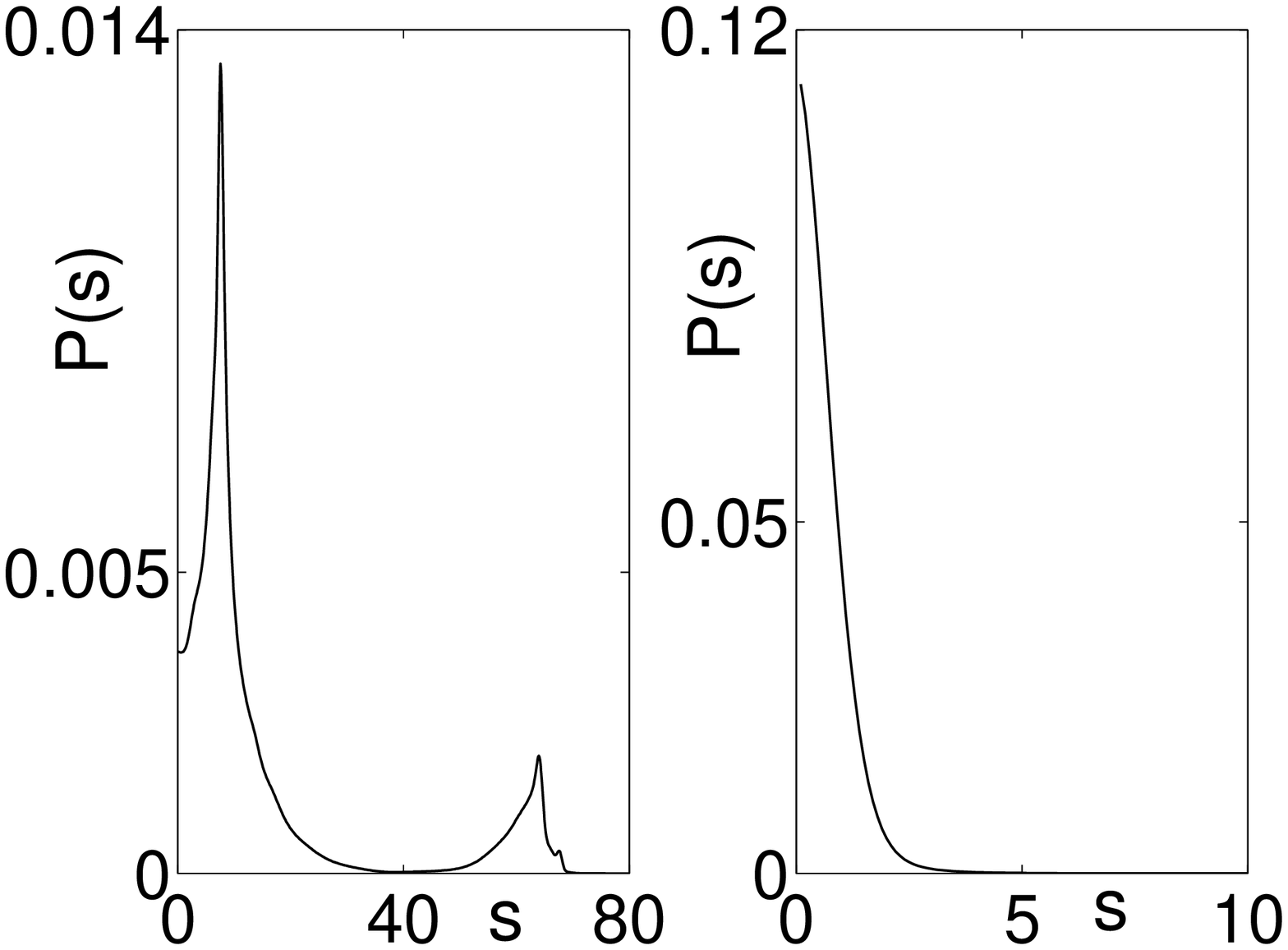}
\begin{figure}
\caption{\label{bigfig8} Distribution of the magnitude of the 
nearest-neighbor height 
difference $s$ for model I with 
$\lambda$=4.0 and $c$=0.02 (left panel), showing the
bimodal nature of the distribution, characteristic 
of a mounded phase with slope selection. The distribution for
$\lambda$=4.0, $c$=0.05 (right panel) does not show this 
behavior.}
\end{figure}}

\vbox{
\epsfxsize=8.0cm
\epsfysize=6.0cm
\epsffile{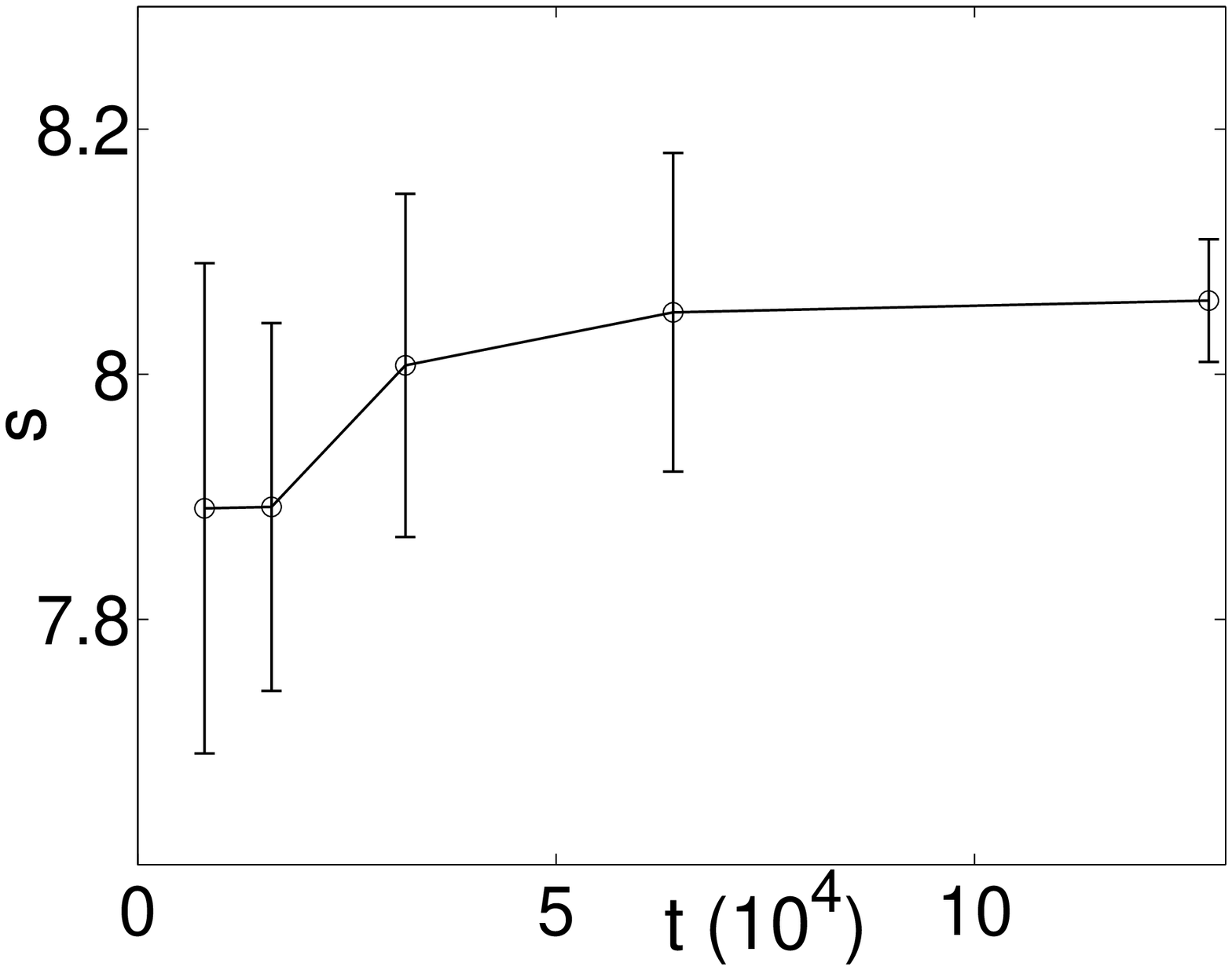}
\begin{figure}
\caption{\label{bigfig9} Average slope of the mounds as a 
function of time for model I in the mounded 
phase ($\lambda=4.0$, $c=0.02$) during the coarsening process, $t$=8000 to 
$t$=$1.28 \times 10^{5}$. The data are for $L$=500 samples 
averaged over 40 runs.}
\end{figure}}

\vbox{
\epsfxsize=8.0cm
\epsfysize=6.0cm
\epsffile{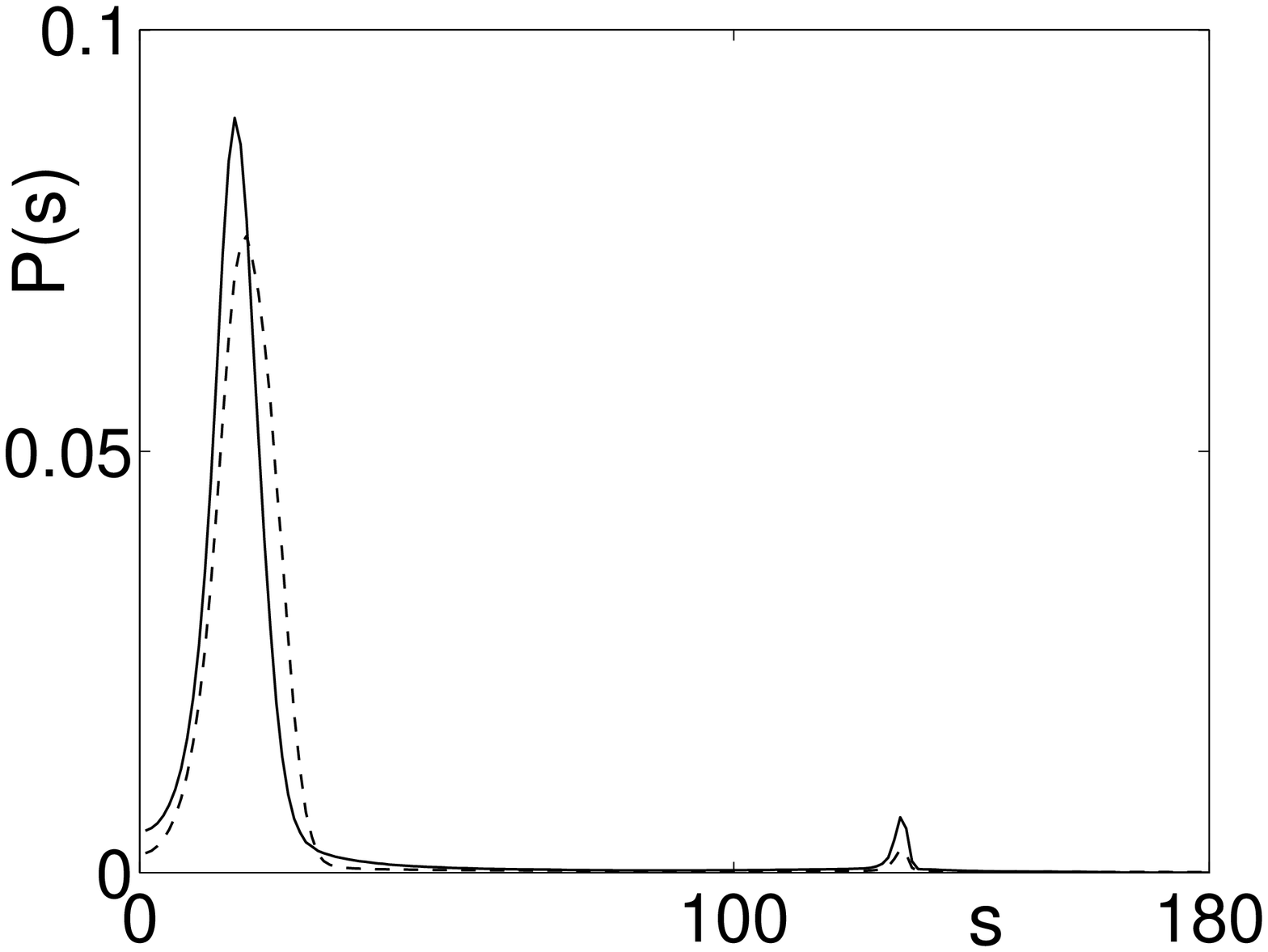}
\begin{figure}
\caption{\label{bigfig10} Distribution of the magnitude of the 
nearest-neighbor height 
difference $s$ for model II with 
$\lambda$=2.0 and $c$=0.005, at two different times, 
$t$=$10^6$ (full line) 
and $t$=$10^7$ (dashed line). 
}
\end{figure}}

\vbox{
\epsfxsize=8.0cm
\epsfysize=6.0cm
\epsffile{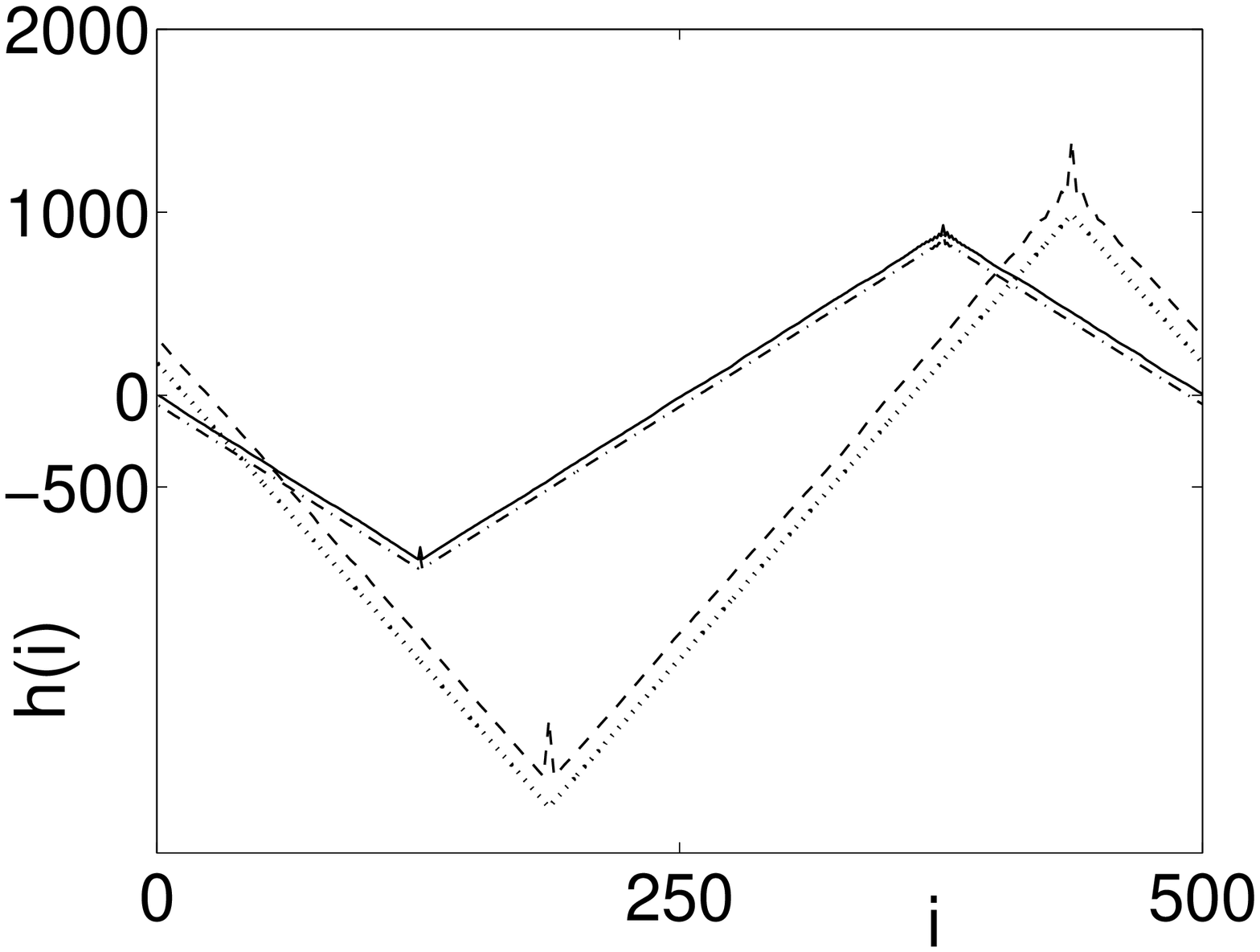}
\begin{figure}
\caption{\label{bigfig11} Fixed point profile for a $L=500$ sample of 
model I with $\lambda$=4.0 and $c$=0.02 (full line), 
compared with a steady state profile (dash-dotted line) 
for the same parameter values. 
The dashed line shows a steady state profile of
a $L=200$ sample of model IA 
with $\lambda=4.0$, $c=0.01$ (both axes scaled by 2.5), and the dotted line
shows an invariant solution of the corresponding continuum equation.}
\end{figure}}

\vbox{
\epsfxsize=8.0cm
\epsfysize=6.0cm
\epsffile{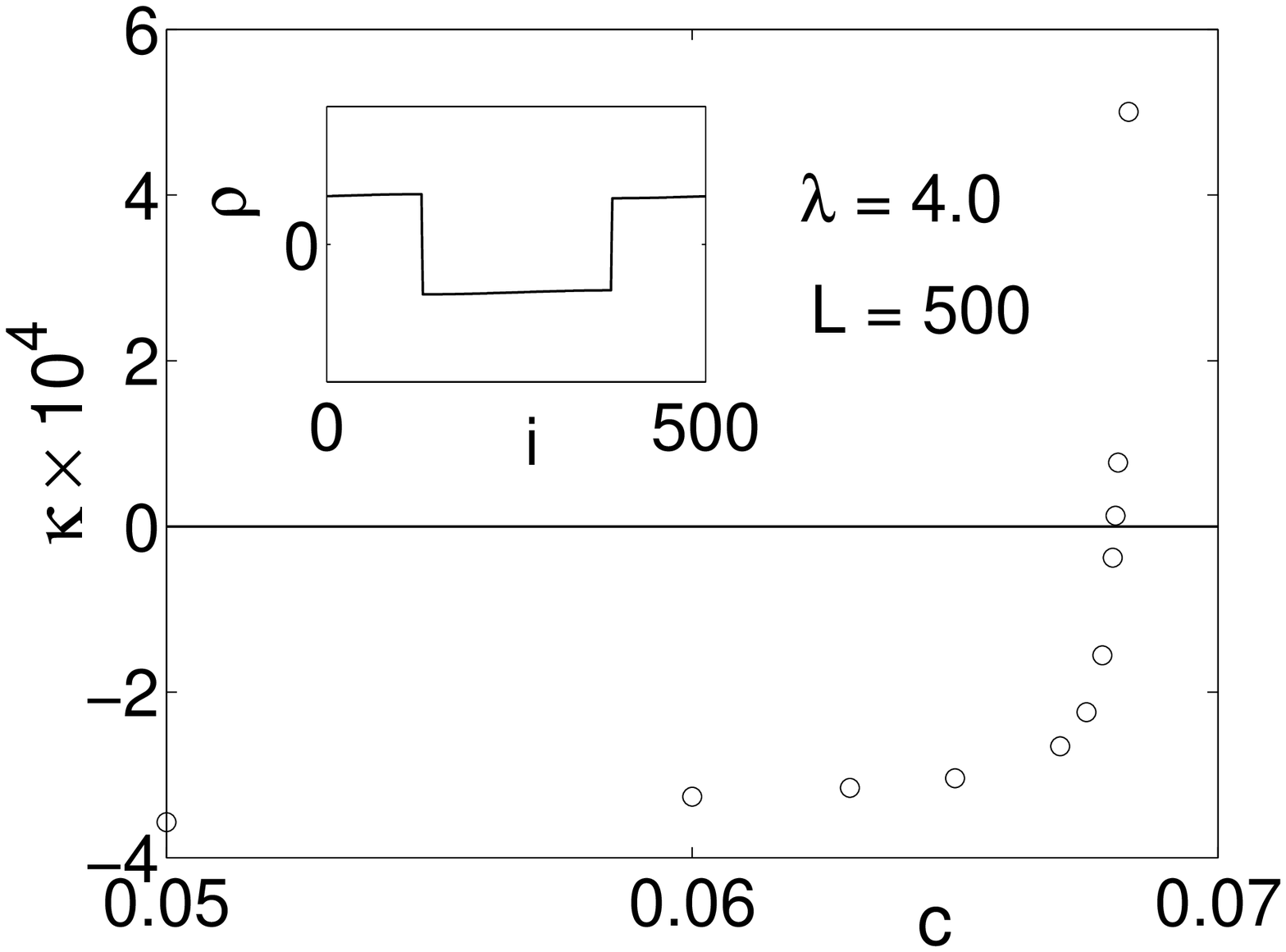}
\begin{figure}
\caption{\label{bigfig12} The dependence of $\kappa$, the closest-to-zero 
eigenvalue of the stability matrix for the mounded fixed point of model I 
with $\lambda$=4.0, $L$=500, on the control parameter $c$. The 
inset shows the right eigenvector $\rho_i$ corresponding to this eigenvalue 
near the point where $\kappa$ crosses zero.}
\end{figure}}

\vbox{
\epsfxsize=8.0cm
\epsfysize=6.0cm
\epsffile{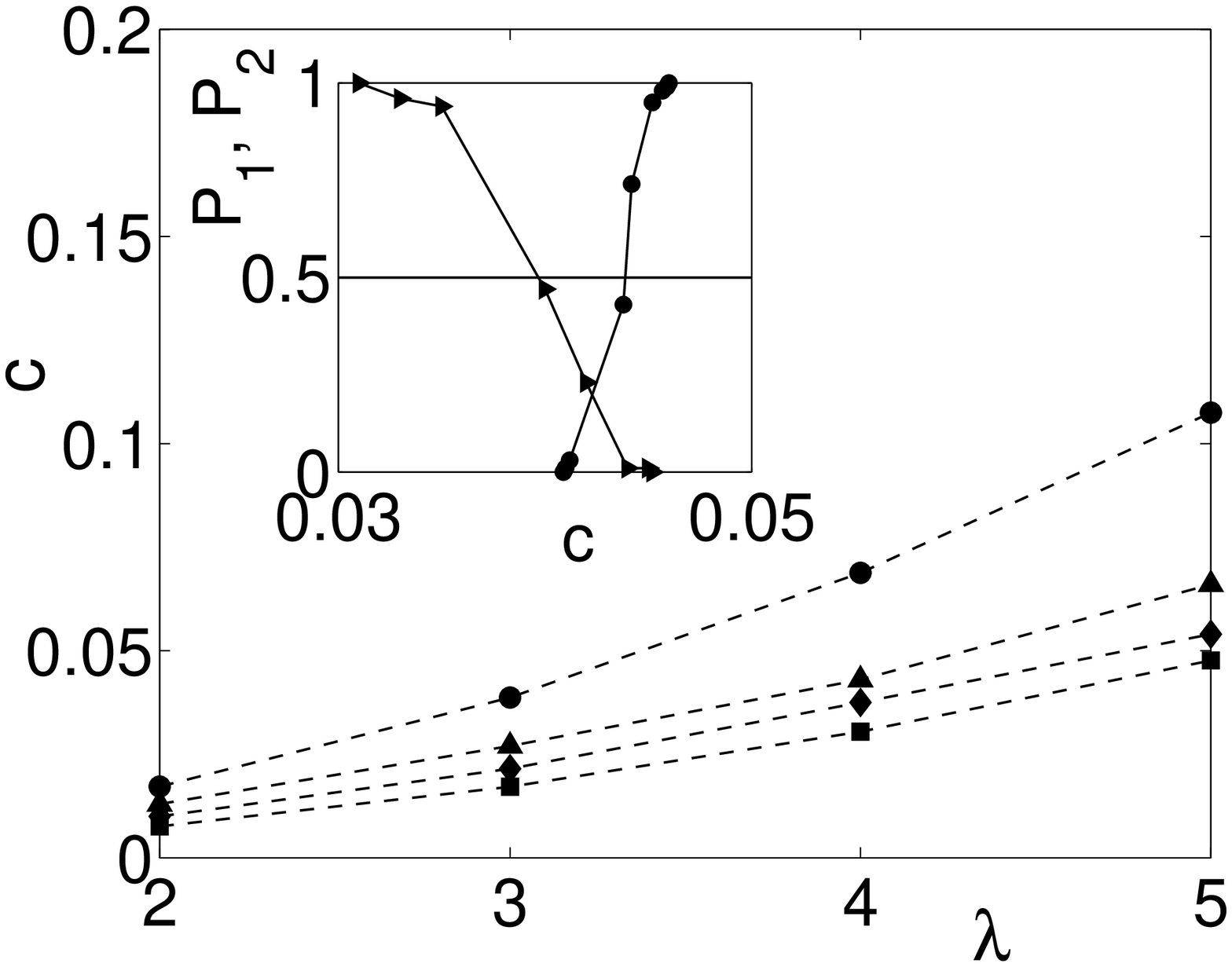}
\begin{figure}
\caption{\label{bigfig13} Critical values of the control 
parameter $c$ as functions of $\lambda$: $c_1$ of model I 
(circles), $c_2$ of model I (triangles), $c_2$ of 
model II (diamonds), and $c_1$ of model IA (squares). Inset: The
probabilities $P_1$ (circles) and $P_2$ (triangles) defined in text, as
functions of $c$ for model I with $\lambda=4.0$, $L=200$.}
\end{figure}}

\vbox{
\epsfxsize=8.0cm
\epsfysize=6.0cm
\epsffile{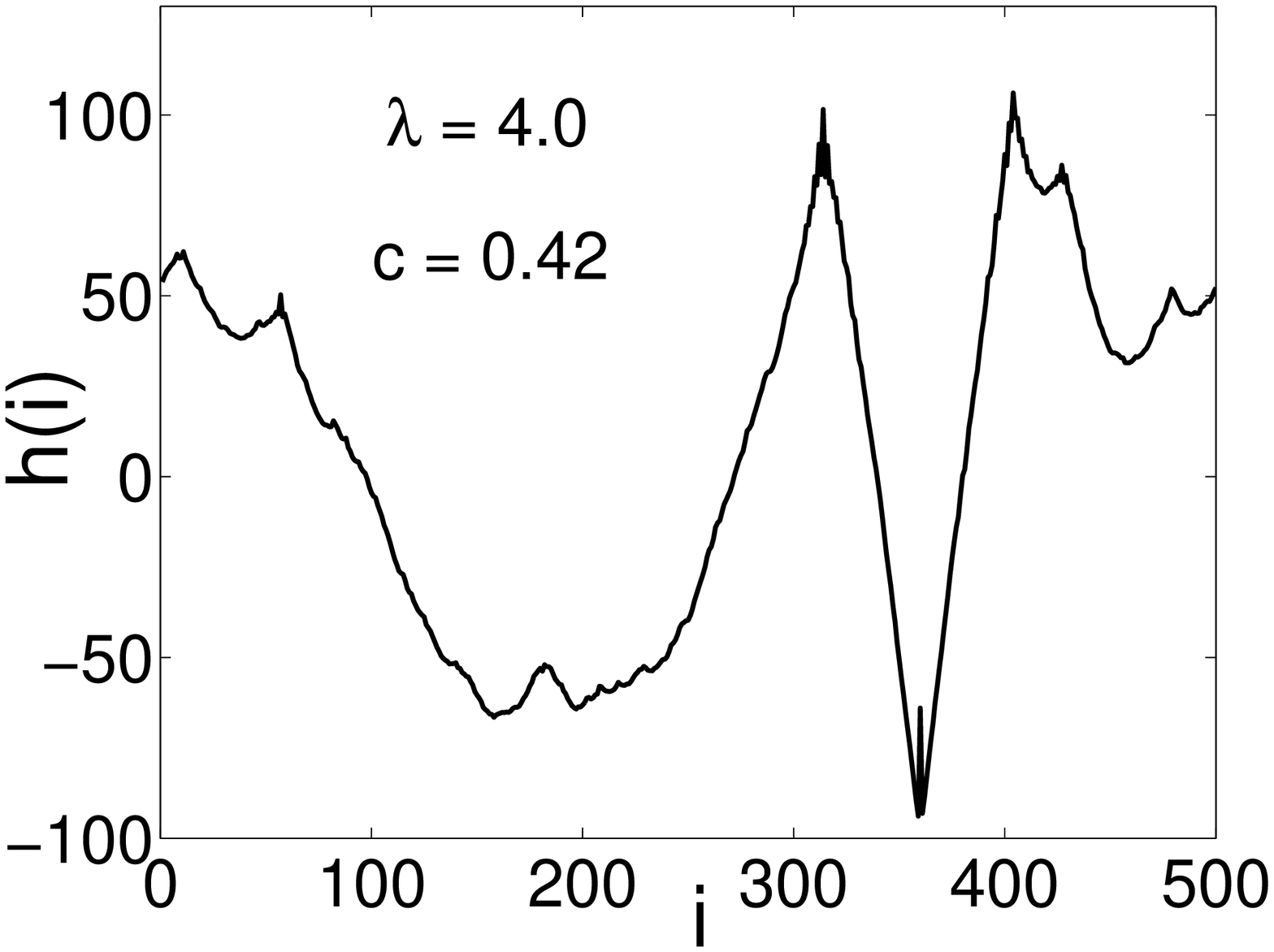}
\begin{figure}
\caption{\label{bigfig14} Two phase coexistence near the phase
transition in model I. The plot shows an interface profile of a
$L=500$ sample with $\lambda$=4.0, $c$=0.42.}
\end{figure}}

\vbox{
\epsfxsize=8.0cm
\epsfysize=6.0cm
\epsffile{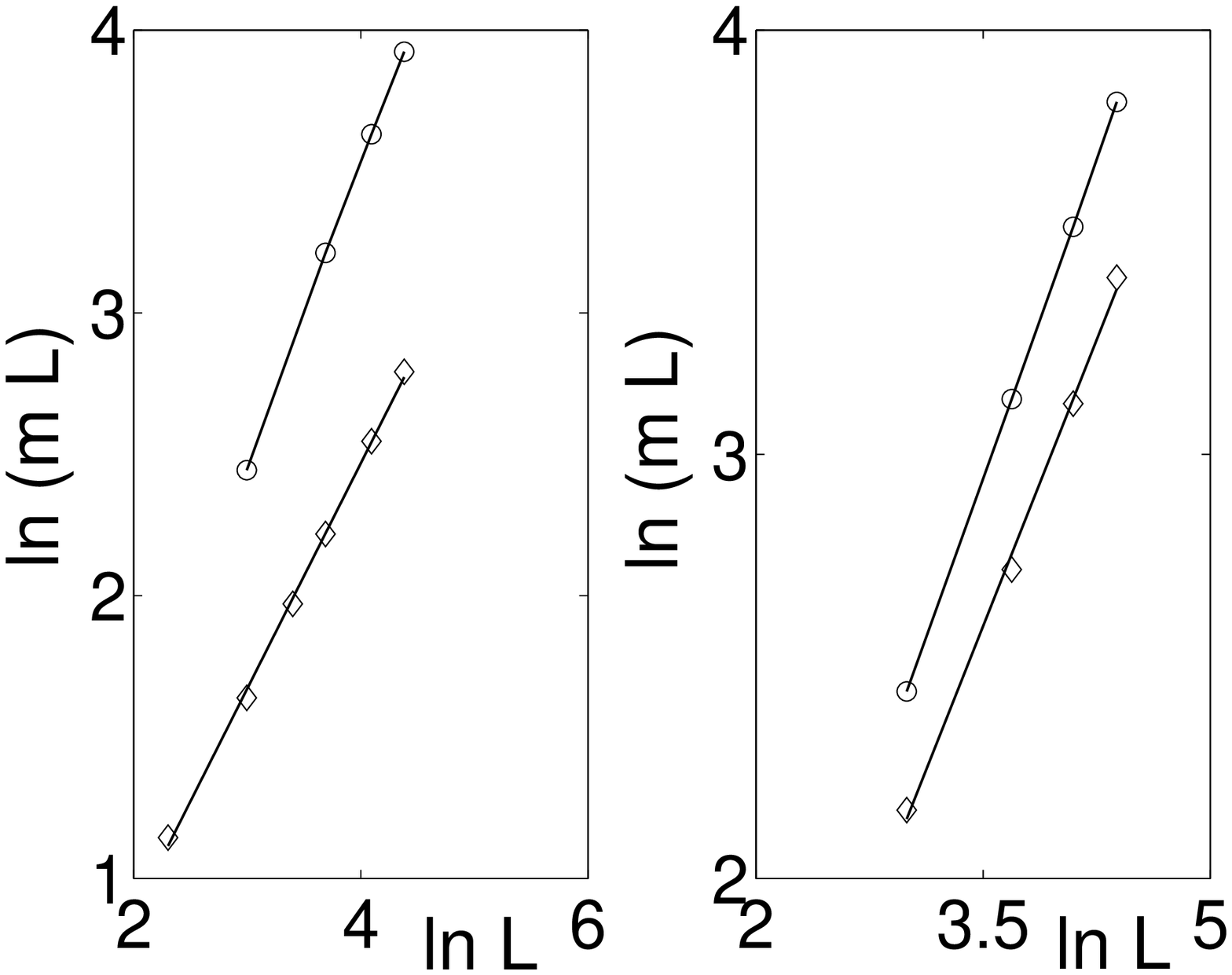}
\begin{figure}
\caption{\label{bigfig15} Finite-size scaling for the 
order parameter $m$. The left panel shows double-log plots of $mL$ as a 
function of the sample size $L$ for model I in the 
mounded phase ($\lambda$=4.0, $c$=0.02, circles) with slope $1.00 \pm 0.01$ 
and in the kinetically rough phase ($\lambda$=4.0, $c$=0.05, diamonds) with 
slope $0.81 \pm 0.02$. The right panel shows similar plots for model II in 
the mounded phase ($\lambda$=2.0, $c$=0.005, circles) with slope 
$1.00 \pm 0.01$ and in the kinetically rough phase ($\lambda$=2.0, $c$=0.015, 
diamonds) $0.88 \pm 0.02$. The straight lines are the best power-law fits 
to the data.}
\end{figure}}

\vbox{
\epsfxsize=8.0cm
\epsfysize=6.0cm
\epsffile{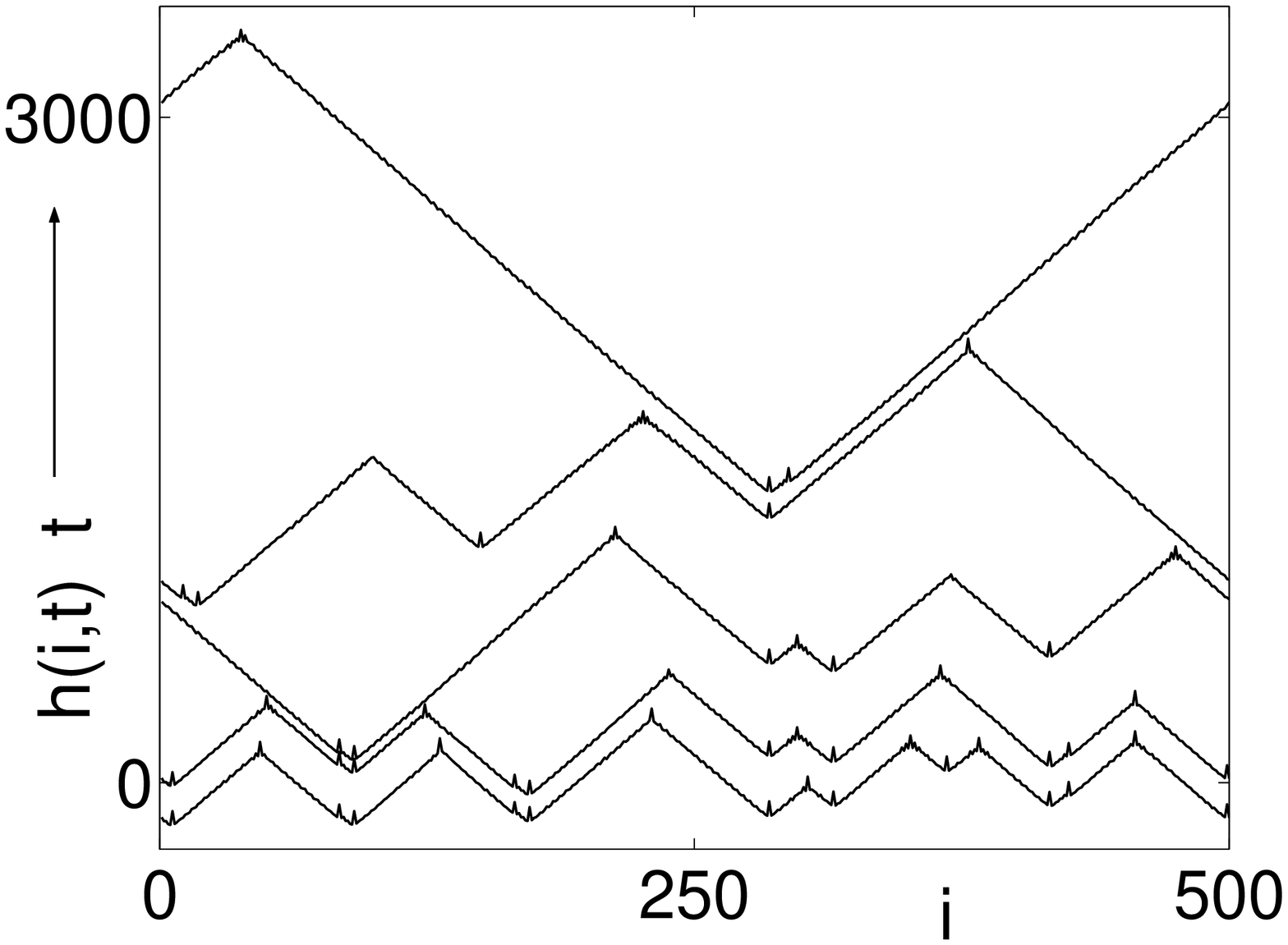}
\begin{figure}
\caption{\label{bigfig16} 
Snapshots ($t$ = $2 \times 10^{4}$, $6 \times 10^{4}$, $10^{5}$, 
$1.4 \times 10^{5}$, and $1.28 \times 10^{7}$) of the profile of a $L=500$ 
sample of model I (profiles at different times have been shifted in the 
vertical direction for clarity) with $\lambda$=4.0, $c$=0.02 in the 
coarsening regime.}
\end{figure}}

\vbox{
\epsfxsize=8.0cm
\epsfysize=6.0cm
\epsffile{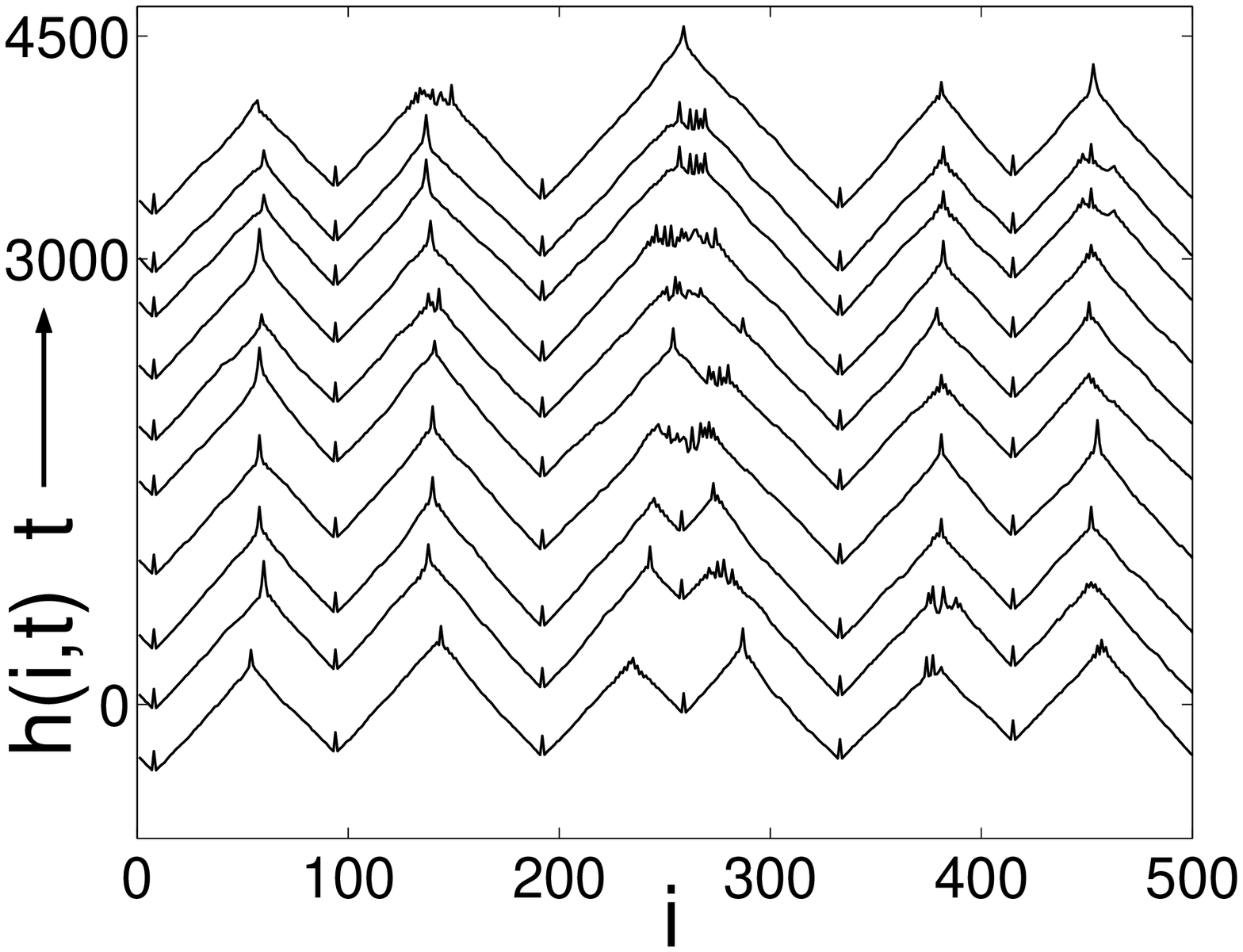}
\begin{figure}
\caption{\label{bigfig17} 
Snapshots ($t$ = $1100000$, $1200000$, $1250000$, $1301000$, $1450000$, 
$1606000$, $1660000$, $1670000$, $1680000$, $1700000$) of the profile 
of a $L=500$ sample of model II (profiles at different times have been 
shifted in the vertical direction for clarity) with $\lambda$=2.0, 
$c$=0.005 in the coarsening regime.}
\end{figure}}

\vbox{
\epsfxsize=8.0cm
\epsfysize=6.0cm
\epsffile{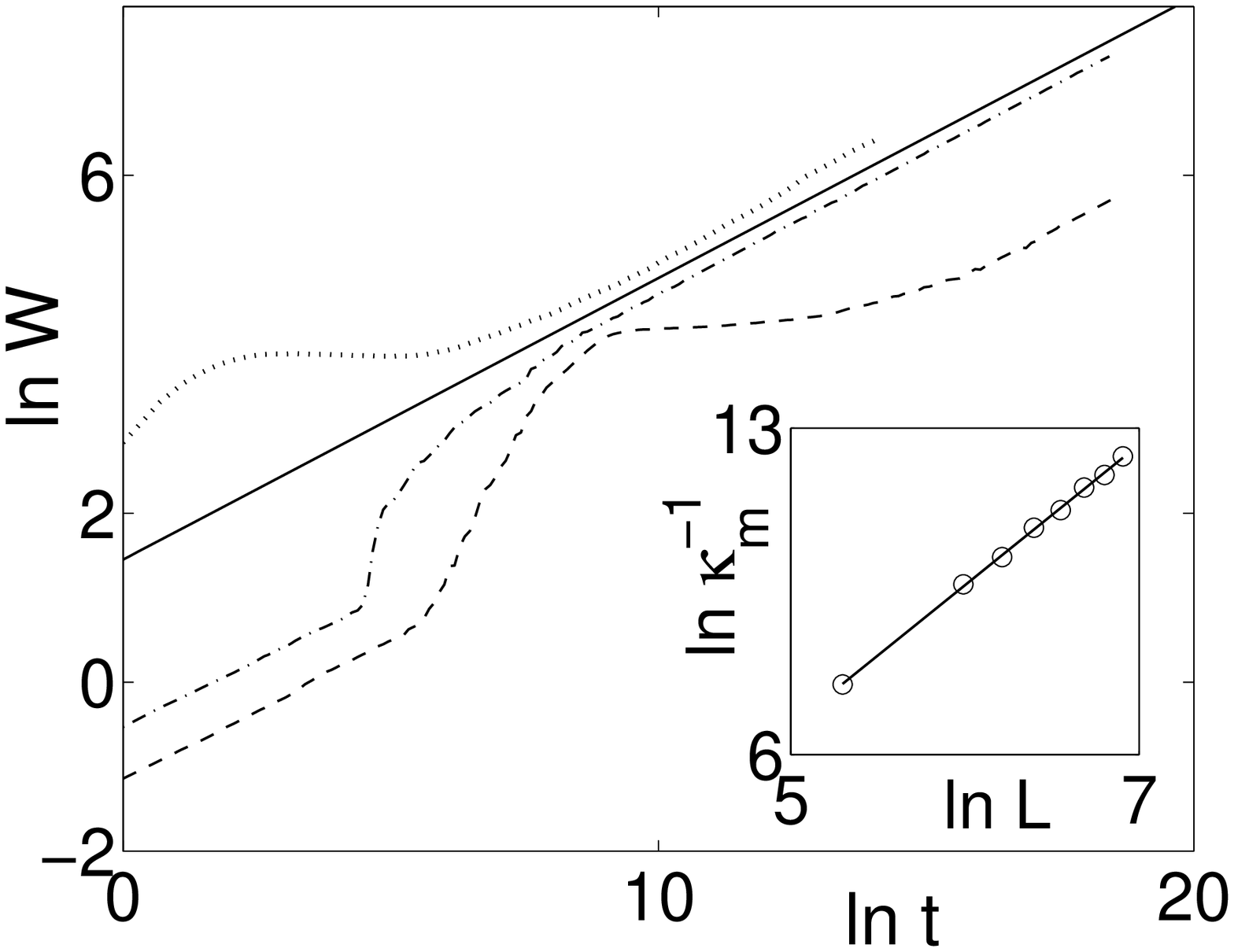}
\begin{figure}
\caption{\label{bigfig18} Double-log plots of the interface 
width $W$ as a function of time $t$ for model I with 
$\lambda$=4.0 and $c$=0.02, with noise (dash-dotted line), 
and without noise (dotted line) averaged over 40 runs for 
$L=1000$ samples. The dashed line shows $W$ versus $t$ data for 
model IA with $\lambda$=4.0, $c$=0.01, averaged over 60 runs 
for $L=500$ samples. The solid line represents power-law behavior with
exponent $n=1/3$. Inset: Finite-size scaling data for the inverse of the
closest-to-zero eigenvalue of the stability matrix for the mounded
fixed point of model I with $\lambda=4.0$, $c=0.02$.}
\end{figure}}

\vbox{
\epsfxsize=8.0cm
\epsfysize=6.0cm
\epsffile{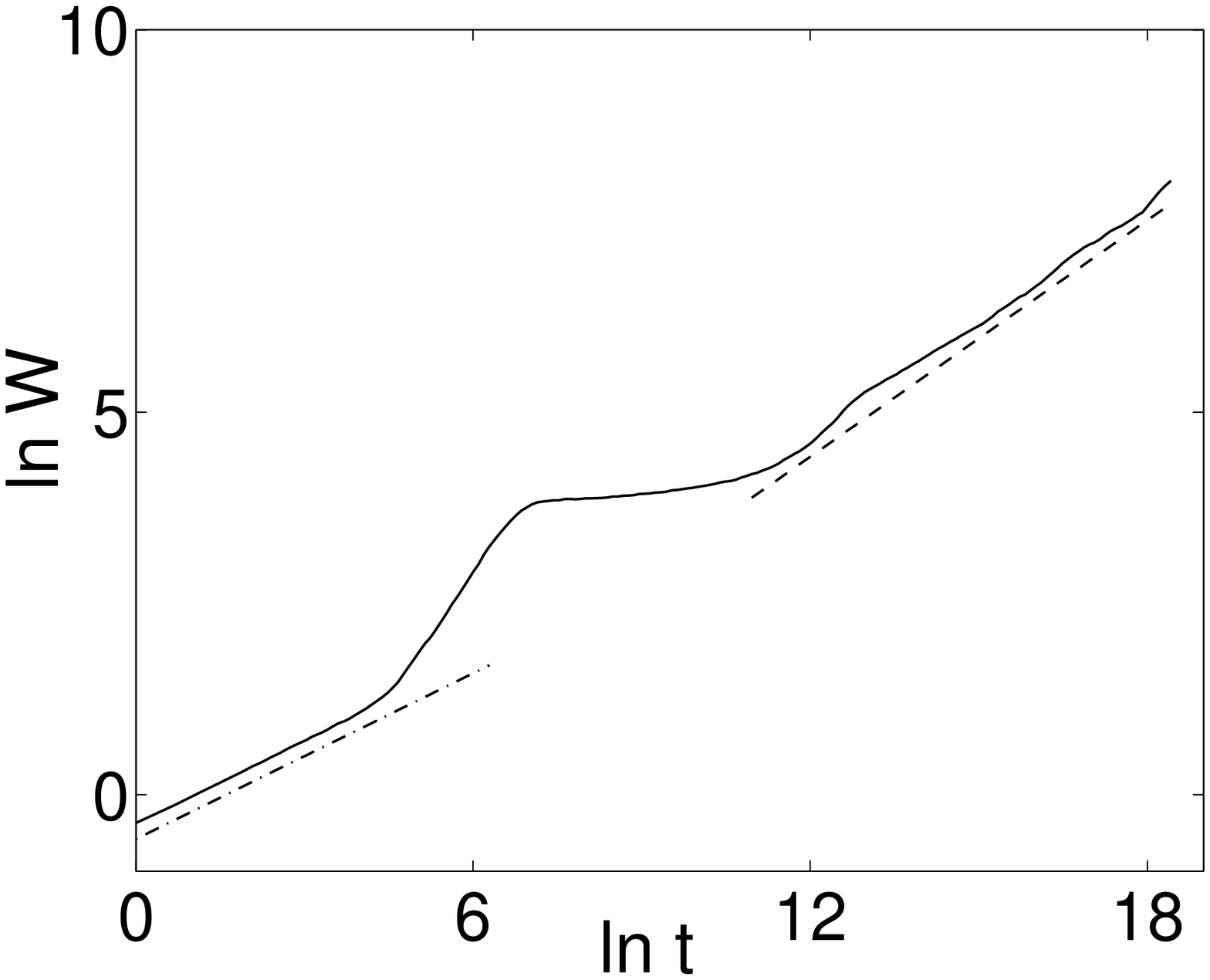}
\begin{figure}
\caption{\label{bigfig19} Double-log plot of the interface 
width $W$ as a function of time $t$ for model II with 
$\lambda$=2.0 and $c$=0.005 (solid line) for $L$=1000 
samples averaged over 40 runs. The dash-dotted and dashed lines
represent power-law behavior with exponent $n=1/3$ and $n=1/2$, respectively.}
\end{figure}}

\vbox{
\epsfxsize=8.0cm
\epsfysize=6.0cm
\epsffile{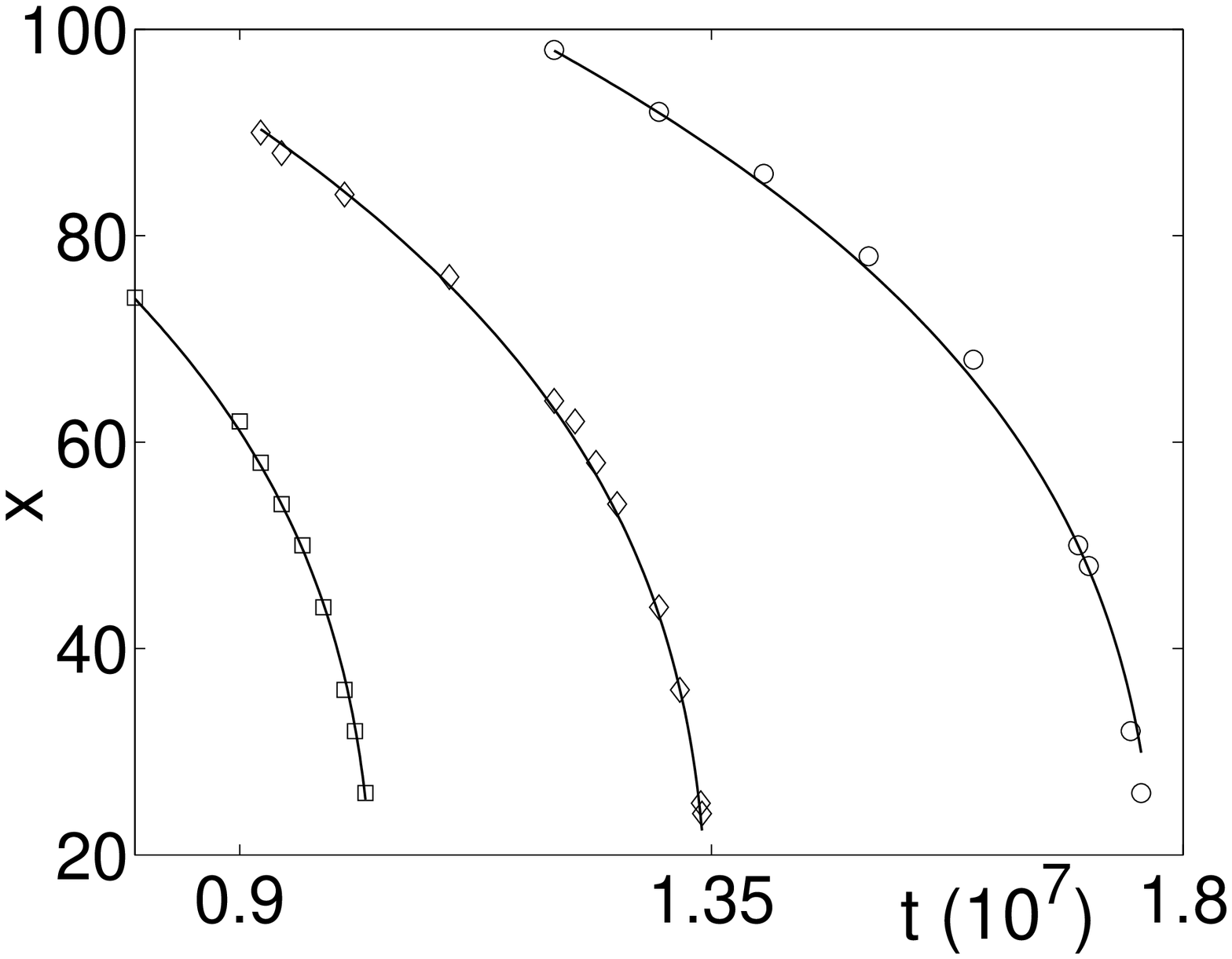}
\begin{figure}
\caption{\label{bigfig20} Peak separation $x$ as a function of time $t$
for a two-mounded structure for model II with parallel updates (see text).
The data shown are for
$\lambda$=2.0, $c$=0.005, $L$=500. The initial value of the
separation is
$x_0$=80 (squares), $x_0$=90 (diamonds), and $x_0$=100 (circles). The
solid lines represent the fits described in the text.}
\end{figure}}

\vbox{
\epsfxsize=8.0cm
\epsfysize=6.0cm
\epsffile{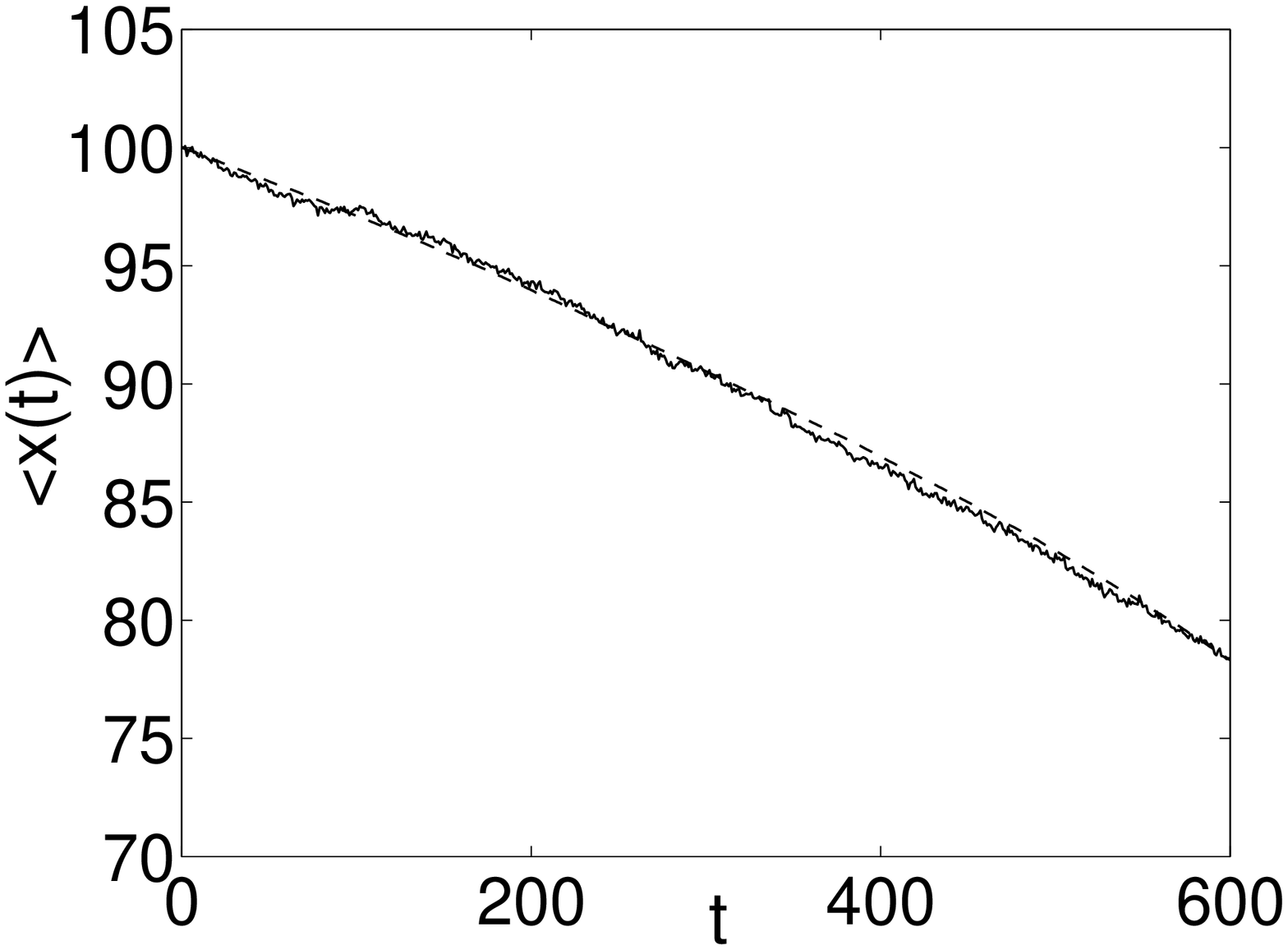}
\begin{figure}
\caption{\label{bigfig21} The solid line shows the 
average value of the separation $x(t)$ between
mound tips (see text) 
as a function of time $t$ for a two-mounded structure for 
model II. The data shown are for $\lambda$=2.0, $c$=0.005, $L$=500, 
$x_{0}$=100, averaged over 800 runs. The dashed line shows 
$\langle x(t) \rangle$ calculated for the reduced model of Eq.(\ref{bw}) with
$C$ = 285.0 and $D$ = 0.15.}
\end{figure}}

\vbox{
\epsfxsize=8.0cm
\epsfysize=6.0cm
\epsffile{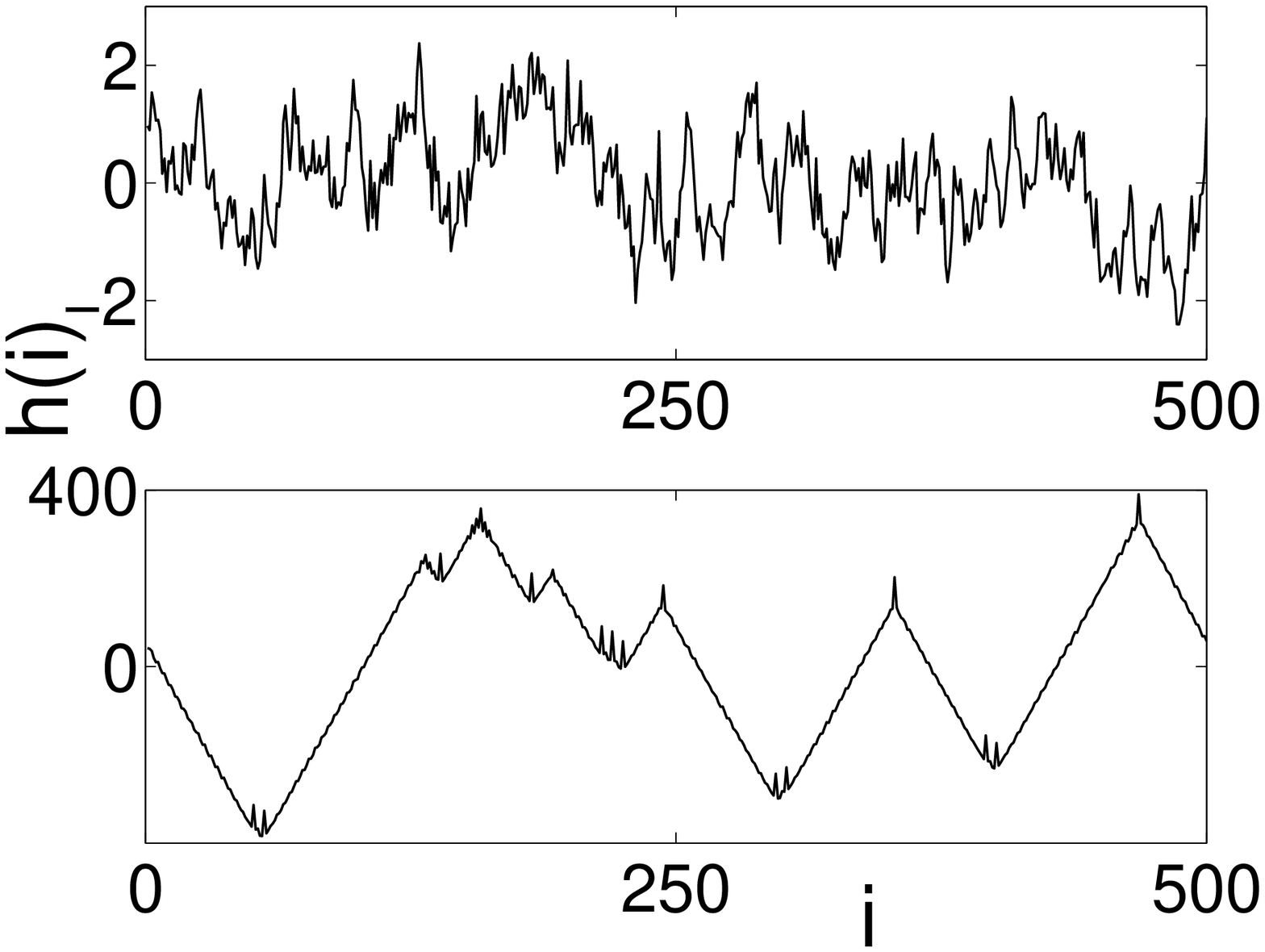}
\begin{figure}
\caption{\label{bigfig22} Profiles at time $t$=$10^5$ for model I with 
conserved noise ($\lambda$=4.0, $c$=0.02, $L$ = 500), for a 
flat initial configuration (top panel), and an initial configuration with
a pillar of height $h_0$ = 1000 (bottom panel).}
\end{figure}}

\vbox{
\epsfxsize=8.0cm
\epsfysize=6.0cm
\epsffile{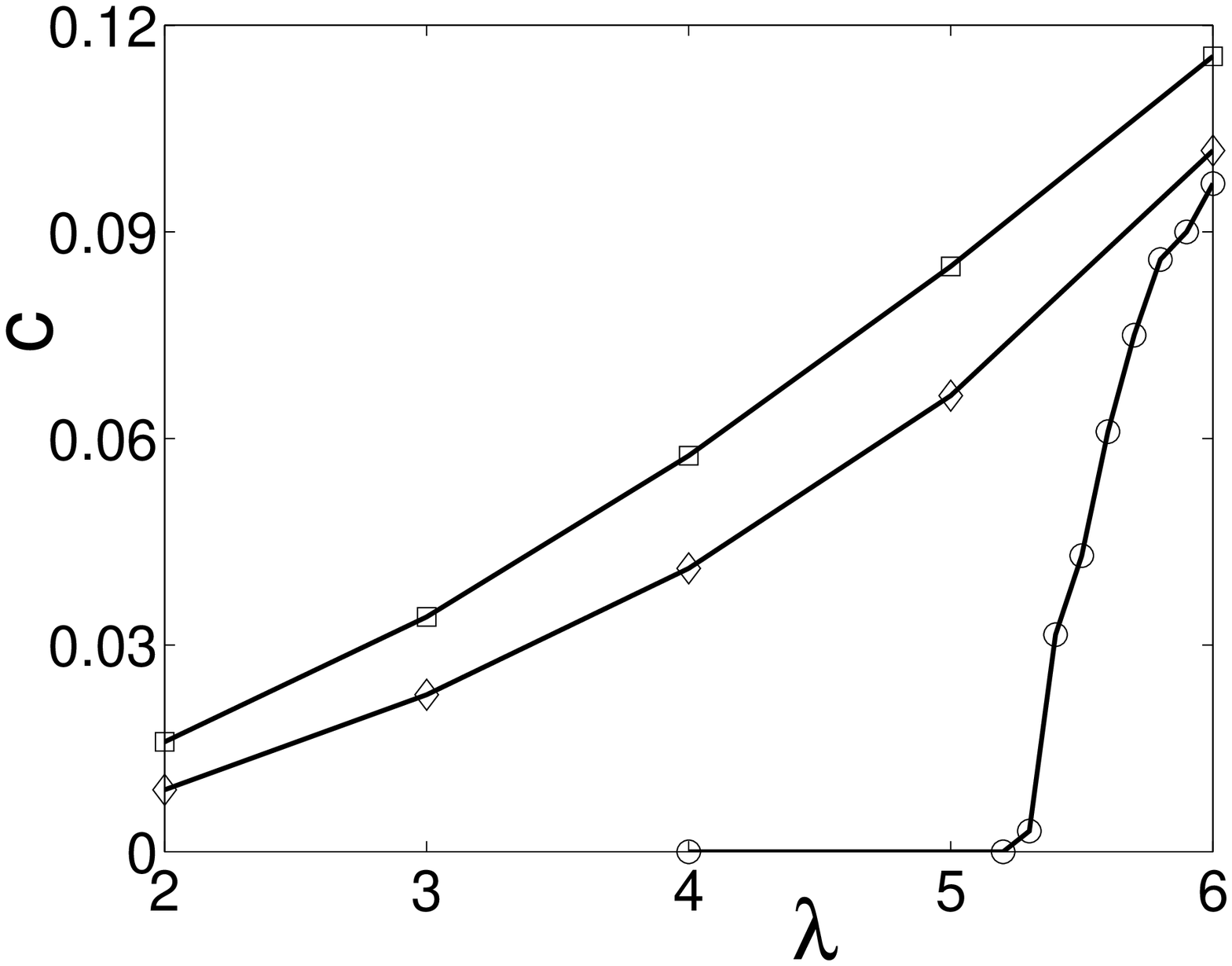}
\begin{figure}
\caption{\label{bigfig23} Phase diagram for model I with 
conserved noise. The plots show $50\%$ stability lines (see text) 
for a flat initial 
state (circles), an initial state with a pillar of height $h_0$ = 1000
(diamonds) and an initial state identical to the mounded fixed point 
of the noiseless equations of motion (squares). The data were 
obtained from 100 $t=10^4$ runs for $L=200$ samples.} 
\end{figure}}
}
\end{multicols}
\end{document}